\documentclass[%
 reprint,
superscriptaddress,
nofootinbib,
 amsmath,amssymb,
 aps,
]{revtex4-2}
\usepackage{mathtools}
\usepackage{aps_macros}
\usepackage{pifont}
\usepackage{graphicx}
\usepackage{dcolumn}
\usepackage{bm}
\usepackage{hyperref}

\usepackage{subfigure}%
\usepackage{float}%
\hypersetup{
     colorlinks   = true,
     citecolor    = blue
}

\hypersetup{colorlinks=true, citecolor=blue, linkcolor = magenta}

\usepackage{tabulary}
\newcolumntype{K}[1]{>{\centering\arraybackslash}p{#1}}

\begin{document}

\preprint{APS/123-QED}

\title{On the Stability of Anisotropic Neutron Stars}

\author{L.~M.~Becerra}
\email{laura.becerra@umayor.cl}
\affiliation{Centro Multidisciplinario de F\'isica, Vicerrector\'ia de Investigaci\'on, Universidad Mayor,  Santiago de Chile 8580745, Chile}
 
\author{E.~A.~Becerra-Vergara}
\email{eduar.becerra@correo.uis.edu.co}
\affiliation{Grupo de Investigaci\'on en Relatividad y Gravitaci\'on, Escuela de F\'isica, Universidad Industrial de Santander A. A. 678, Bucaramanga 680002, Colombia}

\author{F.~D.~Lora-Clavijo} 
\email{fadulora@uis.edu.co}
\affiliation{Grupo de Investigaci\'on en Relatividad y Gravitaci\'on, Escuela de F\'isica, Universidad Industrial de Santander A. A. 678, Bucaramanga 680002, Colombia}

\author{J.~F.~Rodriguez} 
\email{jose.rodriguez2@correo.uis.edu.co}
\affiliation{Grupo de Investigaci\'on en Relatividad y Gravitaci\'on, Escuela de F\'isica, Universidad Industrial de Santander A. A. 678, Bucaramanga 680002, Colombia}

\date{\today}

\begin{abstract}

We model anisotropic neutron stars using three distinct prescriptions for pressure anisotropy—the Horvat, Bowers–Liang, and Covariant models—and three equations of state with different particle compositions, each described by a piecewise-polytropic parametrization with continuous sound speed. The stability of these configurations is assessed through their dynamical evolution using a fully non-linear relativistic code. For stable configurations, we compute the oscillation spectrum and identify the fundamental mode frequency. We found that, while the isotropic and Horvat models become unstable \textcolor{black}{close to} the maximum-mass point, the Bowers–Liang and Covariant models become unstable at lower central densities, indicating that the standard turning-point criterion may not reliably predict the onset of dynamical instability in anisotropic stars.  \textcolor{black}{Based on our results, we also determine the neutral-stability line and verify that configurations lying to the right of this line are indeed unstable under radial perturbations and collapse.}
Overall, given an equation of state, pressure anisotropy can increase the maximum mass of an \textcolor{black}{stable} configuration by up to $\sim 30\%$ compared to the isotropic case. It also allows for more compact stable configurations that may collapse on longer timescales once they become unstable. Finally, we show that these compact stars could initially mimic a black hole's gravitational-wave ringdown. However, the production of subsequent echoes is not guaranteed by high compactness; instead, it depends critically on the star's specific internal structure and equation of state.
\end{abstract}

\maketitle


\section{\label{sec:intro} Introduction}

Neutron stars (NSs) are among the most extensively studied astrophysical objects because they provide a unique environment for exploring physics under extreme density and strong gravity. 
These stars are usually modeled with isotropic pressure distributions \citep{2004Sci...304..536L,2012ARNPS..62..485L,2000ARNPS..50..481H,2016ARA&A..54..401O,2016ApJ...820...28O}, that is, the radial and tangential components of pressure are assumed equal. Theoretical studies indicate that various microphysical phenomena can generate pressure anisotropies within their interiors. These include phase transitions \citep{sokolov1980phase}, pion condensation \citep{1972PhRvL..29..382S}, superfluidity \citep{2000PhR...328..237H,1998NuPhB.531..478C}, and the presence of extremely strong magnetic fields \citep{2012MNRAS.427.3406F,2015MNRAS.447.3278B,2015PhRvD..91d4040F}. These deviations from isotropy can significantly affect the star's structure and macroscopic properties, influencing quantities such as its mass, radius, compactness, moment of inertia, and stability limits \citep{2020EPJC...80..769R,2021ApJ...922..149D,2021EPJC...81..698P,2012PhRvD..85l4023D,2021PhRvC.104f5805R,2022PhRvD.106j3518D,PhysRevD.109.043025,2019PhRvD.100j3006B,2024PhRvD.110j3004B,2022NewAR..9501662K}. 

The fact that anisotropy can significantly modify the internal structure and macroscopic properties of NSs leads us to consider that it could also play a crucial role in their dynamical stability and in the processes driving gravitational collapse. A standard stability criterion for NS requires that the stellar mass, $M$, increases with central density, $\rho_c$: $\partial M/\partial\rho_c>0$ \citep{1966ApJ...145..505B,2011CQGra..28b5009H}.  This approach allows for the determination of the maximum mass that a configuration can sustain before becoming gravitationally unstable and collapsing. A more fundamental stability criterion involves analyzing the frequency of the fundamental mode of radial oscillations \citep{2001A&A...366..565K,haensel2007neutron,glendenning2012compact}. A NS is stable if the squared frequency of this mode is positive ($\omega^2 > 0$), and unstable if it is negative ($\omega^2 < 0$). Recent stability analyses of anisotropic compact stars against radial perturbations \citep{2011CQGra..28b5009H,2012PhRvD..85l4023D,2016JCAP...11..012A,1976A&A....53..283H,2003GReGr..35.1435D,2004CQGra..21.1559K,2017PhRvD..96h3007I,2020EPJC...80..726P,2024PhRvD.109l3039M} show that anisotropy can significantly alter the critical thresholds of compactness and mass. This allows for configurations that sustain higher compactness without violating causality or energy conditions. This indicates that anisotropy may not only delay the collapse into a black hole (BH) but also enable new branches of stable solutions within the stellar parameter space.

These highly compact astrophysical objects can emit gravitational waves (GWs) during their dynamical relaxation phase, known as ringdown, whose signal is practically indistinguishable from that associated with BHs \citep{Cardoso:2016rao}. However, in certain scenarios, the full waveform may include additional structures in the form of recurrent delay pulses, called gravitational echoes. Tentative evidence for these echoes was reported in the post-merger analysis of GW170817 \citep{2017PhRvL.119p1101A}. These signals have been associated with quantum corrections at the horizon scale \citep{2019JCAP...11..010A}.
However, \citet{2018CQGra..35oLT01P} noted that GW echoes could also originate from ultracompact stars possessing a photon sphere.  
In \citep{2019JCAP...04..011U}, it was shown that ultracompact objects constructed with realistic equations of state (EOS) cannot produce GW echoes similar to those from a BH mimic ringdown. In contrast, \citet{2019PhRvD..99j4072R} found anisotropic ultracompact stars (with compactness close to that of a BH) capable of generating GW echoes. However, the physical viability of such configurations remains uncertain, as the analysis did not address whether these models satisfy fundamental criteria such as causality. Since the GWs emission by NS is closely related to their internal structure, analyzing the effect of anisotropy is crucial. In particular, anisotropy could allow achieving levels of compactness suitable for the generation of GW echoes using realistic EOS,  guaranteeing physically consistent configurations.

In this work, we investigate how anisotropy affects the gravitational collapse and dynamical stability of NSs. We use numerical simulations with a fully nonlinear relativistic code that solves the hydrodynamic equations using high-resolution shock-capturing schemes. We also analyze the potential generation of GW echoes in anisotropic NSs, examining the influence of both anisotropy and the star's EOS on this signal. The pressure anisotropy inside the star is modeled using three widely employed prescriptions: the Bowers–Liang model \citep{1974ApJ...188..657B}, in which anisotropy depends nonlinearly on the radial pressure and is induced by gravity; a quasi-local EOS proposed by \citet{2011CQGra..28b5009H}; and a recently developed covariant model by \citet{2019PhRvD..99j4072R}. To explore the interplay between anisotropy and the composition of dense matter, we consider three realistic EOSs (SLy4 \citep{Raduta2015,Chabanat1998,Danielewicz2009}, GM1Y6 \citep{Glendenning1991,Oertel2015}, and QHC21 \citep{Kojo2022}) representing NS matter with nucleons, hyperons, and quarks, respectively. These EOSs are implemented via a piecewise-polytropic parametrization that ensures continuity of the radial pressure and sound speed throughout the star. 

The paper structure is as follows. In Sec.~\ref{sec:NEM} we outline the formalism for constructing anisotropic stellar configurations, including the anisotropy models Sec.~\ref{subsec:anisotropy}, the EOS parameterization via piecewise polytropic fit Sec.~\ref{subsec:EOS}, and a brief description of the numerical code Sec.~\ref{subsec:code}. Sec.~\ref{sec:SC} presents the computation of the fundamental radial oscillation mode through perturbative evolution of stable configurations. We study the gravitational collapse of anisotropic NSs in Sec.~\ref{sec:collapse}. The analysis of the ringdown signal and the impact of anisotropy and the EOS are discussed in Sec.~\ref{sec:ecos}. Finally, we give our concluding remarks in Sec.~\ref{sec:discussion}. Throughout this paper, we use geometrical units where $c = G = 1$, unless stated otherwise, and adopt the metric signature convention $(-, +, +, +)$.

\section{\label{sec:NEM} Basic Equations}

The dynamics of relativistic fluids in curved spacetime are governed by the general relativistic fluid equations, which are derived from the following basic equations:
\begin{itemize}
    \item Einstein's Field Equations
    \begin{equation}
        G_{\alpha \beta} = 8 \pi T_{\alpha \beta}, \label{eq:EFE}
    \end{equation}
    where $G_{\alpha \beta}$ is the Einstein tensor and $T_{\alpha \beta}$ the total stress-energy tensor.
    \item Conservation of the baryon number
    \begin{equation}
        \nabla_\alpha (\rho u^\alpha)=0, \label{eq:CBN}
    \end{equation}
    where $\nabla_\alpha$ is the covariant derivative consistent with the metric tensor $g_{\alpha \beta}$, $u^\alpha$ is the the fluid four velocity, and $\rho$ is the fluid rest-mass density.
    \item Conservation of the energy-momentum 
    \begin{equation}
        \nabla_\alpha T^{\alpha\beta}=0 . \label{eq:CEM}
    \end{equation}
\end{itemize}
We assume matter is described by an anisotropic fluid with the energy-momentum tensor \cite{misner1973,2016Pimentel}:
\begin{equation}\label{eq:Tmunu}
T_{\alpha \beta} = (\epsilon + P_\perp) u_\alpha u_\beta + P_\perp g_{\alpha \beta} + (P - P_\perp) k_\alpha k_\beta,
\end{equation}
where $P = P(r,t)$ is the radial pressure, $P_\perp = P_\perp(r,t)$ is the tangential pressure, and $\epsilon = \epsilon(r,t)$ is the energy density. The fluid 4-velocity $u^\alpha = [u^t, u^r, 0, 0]$ satisfies $u^\alpha u_\alpha = -1$. The spacelike vector $k^\alpha = [k^t, k^r, 0, 0]$ is normalized ($k^\alpha k_\alpha = 1$) and orthogonal to $u^\alpha$ ($u^\alpha k_\alpha = 0$).

We consider a spherically symmetric spacetime described by the following line element in spherical coordinates:
\begin{equation}
\mathrm{d}s^2 = -\alpha^2 \mathrm{d}\mathrm{t}^{2} + a^2 \mathrm{d}r^2  + r^2\mathrm{d}\theta+r^2\sin^2\theta\mathrm{d}\varphi^2,
\label{eq:ds2}
\end{equation}
where $\alpha=\alpha(r,t)$ and $a=a(r,t)$ are metric functions depending on the radial coordinate $r$ and time $t$. 
Using the above definitions, the components of the four-velocity $u^\alpha$ and an orthogonal spacelike vector $k^\alpha$ are:
\begin{align}
u^t &= \frac{W}{\alpha}, & u^r &= W v^r, \\
\label{eq:k_vec} k^t &= \frac{\sqrt{1-W^2}}{\alpha}, & k^r &= \frac{W}{a},
\end{align}
being $W = (1 - v_r v^r)^{-1/2}$ the Lorentz factor
, and $v^r$ the radial component of the spatial velocity.

Using the line element (\ref{eq:ds2}), the system of equations (\ref{eq:CBN})-(\ref{eq:CEM}) can be expressed in Valencia's formulation \cite{1991PhRvD..43.3794M,1997ApJ...476..221B} as:
\begin{equation}\label{eq:tovc}
    \partial_t (a \mathbf{q}) + \frac{1}{r^2} \partial_r \left[ \alpha a r^2 \mathbf{f}(\mathbf{q}) \right] = \mathbf{s}(\mathbf{q}),
\end{equation}
where:
\begin{itemize}
    \item $\mathbf{q} = [D, S_r, \tau]^\top$ is the vector of conservative variables
    \item $\mathbf{f}(\mathbf{q}) = [D v^r, S_r v^r + P, (\tau + P) v^r]^\top$ is the radial flux vector
    \item $\mathbf{s}(\mathbf{q}) = [0, S_M/r^2, -S_E/r^2]^\top$ is the source vector
\end{itemize}
The conservative variables are defined in terms of the primitive variables  $\{\rho, P, v_r\}$,  as:
\begin{align}
    D &= \rho W, \\
    S_r &= \rho h W^2 v_r, \\
    \tau &= \rho h W^2 - P - \rho W,
\end{align}
where $h = (\epsilon + P)/\rho$ is the specific enthalpy. The source terms are given by:
\begin{align}
    S_E &= a \alpha m S_r, \\
    S_M &= a \alpha \left[ 2r (P - \sigma) - m a^2 (P + S_r v^r + \tau + D) \right], 
\end{align}
with $m = r (1/a^2 - 1)/2$ being the mass function, and to characterize the transition between isotropic and anisotropic matter configurations, we introduce the anisotropy parameter: $\sigma \equiv P - P_\perp$.

Finally, the Einstein field equations (\ref{eq:EFE}) for the line element (\ref{eq:ds2}) take the following form:
\begin{align}
    \frac{\partial_r a}{a^3}  &= -\frac{m}{r^2} + 4\pi r (\tau + D), \label{eq:Einstein1} \\
    \frac{\partial_t a}{a}  &= -4\pi r \alpha S_r, \label{eq:Einstein2} \\
    \frac{\partial_r \alpha}{\alpha}  &= a^2 \left( \frac{m}{r^2} + 4\pi r P + 4\pi r S_r v^r \right). \label{eq:Einstein3}
\end{align}
The equations governing the conservative variables $D$, $S_r$, and $\tau$, together with the metric functions $a$ and $\alpha$, are given by equations~(\ref{eq:tovc}), (\ref{eq:Einstein2}), and (\ref{eq:Einstein3}). Equation~(\ref{eq:Einstein1}) corresponds to the Hamiltonian constraint. The system is fully determined once the anisotropy model and the fluid EOS are specified.

\subsection{Anisotropy models}\label{subsec:anisotropy}
 We examine three fundamental anisotropy models from the literature: the Horvart model \citep{PhysRevD.85.124023,PhysRevD.109.043025}, the Bowers-Liang model \citep{1974ApJ...188..657B}, and the Covariant model \citep{2019PhRvD..99j4072R}. Their respective functional forms are given by:
\begin{eqnarray}
\sigma_H    &=&  -\lambda_H  \left(\frac{2 m}{r}\right)^2 P \, , \label{eq:Delta_p}\\
\sigma_{BL} &=& -\lambda_{BL} (\epsilon + 3P)(\epsilon + P) \left( 1-\frac{2  m }{r} \right)^{-1} r^{2}, \label{eq:sigma_BL}\\
\sigma_{C} &=& \lambda_{C} f(\epsilon)k^{\nu} \nabla_\nu P \, ,\label{eq:sigma_C}
\end{eqnarray}
where $\lambda_H$, $\lambda_{BL}$, and $\lambda_C$ are  constant parameters controlling the degree of anisotropy inside the star.

For the Horvat model, we have adopted the formulation presented in \cite{2024PhRvD.110h3020L}, which ensures regular conditions at the star’s center under non-radial perturbations.  To guarantee causality for the radial and tangential sound speeds, we constrain the anisotropy parameter for the Horvat model to the range $-3.5 \leq \lambda_H \leq 3.5$.

Following \cite{2025PhRvD.111j3005B}, the anisotropy parameter for the Bowers–Liang model is constrained to the range $0 \leq \lambda_{BL} < 1$, to ensure that the radial pressure is always a decreasing function of radius and that the tangential sound speed remains finite at the star's center.

Finally, for the covariant model, we use \cite{2025PhRvD.111j3005B}:
\begin{equation}
f(\epsilon)=\epsilon_c-\epsilon    
\end{equation}
Then, for the line element given in equation~(\ref{eq:ds2}) and the space-like vector of equation~(\ref{eq:k_vec}), the covariant anisotropy model takes the form:
\begin{equation}\label{eq:simgac_e}
    \sigma_C = \lambda_C (\epsilon_c-\epsilon)\left[\frac{v_r}{\alpha}\frac{\partial P}{\partial t} +\frac{\partial P}{\partial r} \right]\frac{W}{a}\, .
\end{equation}
The anisotropy parameter is constrained to the range: $0\leq \lambda_C\leq 4.4 \times 10^{3} M_\odot^3 $.

 \subsection{Equation of state}\label{subsec:EOS}

\begin{table*}
    \begin{tabular}{c|cccc|ccccc|c}\hline
     &  &  &  &  &  &   & & & \\
        $\quad$ \textbf{EOS}  &  $\mathbf{\log \rho_1} $  & $ \mathbf{\log \rho_2} $  &  $ \mathbf{\log \rho_3} $  &$ \mathbf{\log \rho_4} $    & $\Gamma_1$ & $\Gamma_2$ & $\Gamma_3$ & $\Gamma_4$& $\Gamma_5$ & $K_5$ \\
           &$[\mathrm{g\, cm^{-3}}]$ &  $[\mathrm{g\, cm^{-3}}]$  &$[\mathrm{g\, cm^{-3}}]$ &  $[\mathrm{g\, cm^{-3}}]$ &  &  & & & & [cgs units]\\
         &  &  & & &  &  &   & & & \\\hline
        &  &  &  &  &  &   & & & \\
        SLy4 \cite{Raduta2015,Chabanat1998,Danielewicz2009} & $14.99$ &  $14.87$ &  $13.90$ & $12.84$ & $2.68$  &   $2.66$   & $3.02$  &  $1.38$ & $1.29$ &$2.77\times10^{14}$\\
        GM1Y6 \cite{Glendenning1991,Oertel2015} & $14.99$ &  $14.87$ & $13.76$  & $12.84$ & $1.64$ &$1.63$ &$2.96$ & $1.38$ & $1.29$ &  $2.78\times10^{14}$ \\
        QHC21 \cite{Kojo2022}  & $14.99$ &  $14.87$ & $13.99$  & $12.84$ & $2.22$ &$2.11$ &$3.26$ & $1.38$ & $1.29$ &  $2.78\times10^{14}$\\
       
         &  &  &  &  &  &   & & & \\\hline
        \hline
    \end{tabular}
    \caption{ Generalized Piecewise Polytropic  fit parameters.}   \label{tab:EoS}
\end{table*}

The EOS defines the relationship between energy density, $\epsilon$, and the pressure, $P$. For this work, we use NS EOSs that can model objects with more than $2~M_\odot$, in agreement with astrophysical observations \cite{2019ApJ...887L..21R,2022ApJ...934L..17R}.   From the available EOSs in the literature, we selected SLy4, GM1Y6, and QHC21, which describe matter composed of nucleons; nucleons and hyperons; and nucleons, hyperons, and quarks, respectively. 

To simplify the numerical implementation, we parametrize these EOSs using the Generalized Piecewise Polytropic (GPP) fit \cite{Boyle2020}. We divide the baryonic rest-mass density range in $N$ intervals.  Within each interval, spanning from $\rho_i$ to $\rho_{i+1}$, the pressure and energy density are given by:
\begin{eqnarray}\label{eq:poly_eos}
    P(\rho) &=& K_i\rho^{\Gamma_i} + \Lambda_i \,,\\
    \epsilon(\rho) &=& \frac{K_i}{\Gamma_i-1}\rho^{\Gamma_i} +(1+a_i)\rho - \Lambda_i \,, 
\end{eqnarray}
where the fit parameters are the polytropic indices $\Gamma_i$ and the dividing densities $\rho_i$. The constants $K_i$, $\Lambda_i$, and $a_i$ are determined by ensuring the continuity of the energy density, pressure, and sound speed at the dividing densities:
\begin{eqnarray}
    K_{i+1}&=& K_i\frac{\Gamma_i}{\Gamma_{i+1}}\rho^{\Gamma_i-\Gamma_{i+1}} \,,\label{eq:Ki}\\
    \Lambda_{i+1}&=& \Lambda_i + \left(1-\frac{\Gamma_i}{\Gamma_{i+1}}\right)K_i\rho^{\Gamma_i} \,,\label{eq:Lambdai}\\
    a_{i+1}&=& a_i +\Gamma_i\frac{\Gamma_{i+1}-\Gamma_i}{(\Gamma_{i+1}-1)(\Gamma_i-1)}K_i\rho^{\Gamma_i-1} \,.\label{eq:ai}
\end{eqnarray}
 
For the high-density region of the EOSs (densities greater than the nuclear saturation density, $\rho_s\approx 2.4 \times 10^{14}$ g cm$^{-3}$), we use a three-zone GPP model, while for the low-density region ($\rho<\rho_s$), we adopt a two-zone parameterization  (see Table~\ref{tab:EoS}; for more details, see \cite{PhysRevD.109.043025}). The above recurrence relations are solved by setting the parameters in the lowest density zone to  $a_5=0$, $\Lambda_5=0$ in cgs units (  for $K_5$ see Table \ref{tab:EoS} ). 

\subsection{Code and Initial set-up}\label{subsec:code}

We adapted the code from \cite{2012arXiv1212.1421G} to evolve anisotropic stellar configurations with a piecewise EOS. The code employs a high-resolution shock-capturing method utilizing the Harten, Lax, and Leer (HLLE) Riemann solver and the minmod linear piecewise reconstructor. For time integration, we used a third-order Runge–Kutta TVD integrator. At each time step, the primitive variables are recovered from the conservative variables using a Newton-Raphson routine.

The simulations use a one-dimensional spherical grid from $r = 0$ to $r = 80 \, M_\odot$. All runs employ a uniform spatial resolution of $\Delta r = 5.1\times 10^{-3} \, M_\odot$ (a convergence test is shown in Appendix~\ref{app:I}). Outflow boundary conditions are applied to the conservative variables, and the metric functions are extrapolated.

In particular, implementing the Horvat and Bowers–Liang anisotropic models in the code only requires modifying the source term in equation~(\ref{eq:tovc}), as their functional forms are expressed directly in terms of the primitive variables. In contrast, implementing the covariant anisotropy model also requires computing the time derivative of the radial pressure (see equation~\ref{eq:simgac_e}). Since the radial pressure can be written as
\begin{equation}
    P = \frac{\tau + D + P}{W^2} - \epsilon \, .
\end{equation}
Its time derivative, after some algebraic manipulation, takes the form:
\begin{eqnarray}
     \left(\frac{\partial \epsilon}{\partial P}- v_rv^r\right)   \frac{\partial P}{\partial t} &=&\left(1+v_rv^r\right)\frac{\partial(\tau + D)}{\partial t} \nonumber \\  
     & & -2 v^r\left(\frac{\partial S_r}{\partial t} - \frac{S_r}{a}\frac{\partial a}{\partial t} \right). 
\end{eqnarray}
For the initial setup, we build spherically anisotropic configurations by solving the TOV equations:
\begin{eqnarray} 
\dfrac{\mathrm{d}m}{\mathrm{d}r} &=& 4\pi r^{2}\epsilon , \label{eq:TOV_a}\\
\dfrac{\mathrm{d}P}{\mathrm{d}r} &=& -\dfrac{\left(\epsilon + P\right)\left( m + 4\pi r^3 P\right)}{r\left(r-2m\right)} - \frac{2\sigma}{r}, \label{hyd}\\
\dfrac{1}{\alpha}\dfrac{\mathrm{d}\alpha}{\mathrm{d}r} &=& \dfrac{ m + 4\pi r^3 P}{r\left(r-2m\right)}\label{eq:TOV_b},
\end{eqnarray}
using a fourth-order Runge-Kutta integrator. We then perturb the rest-mass density according to:
\begin{eqnarray}
   \frac{ \delta \rho}{\rho} = A\sin(\pi r/R_{\rm star}) \, ,
\end{eqnarray}
where   $A$ is the perturbation amplitude and $R_{\rm star}$ is the coordinate point at which the rest-mass density drops to about $\rho_{\rm atm}\sim10^{7}$ g\, cm$^{-3}$.

\section{\label{sec:SC} Stable configurations}

\begin{figure}
    \centering
    \includegraphics[width=0.95\linewidth]{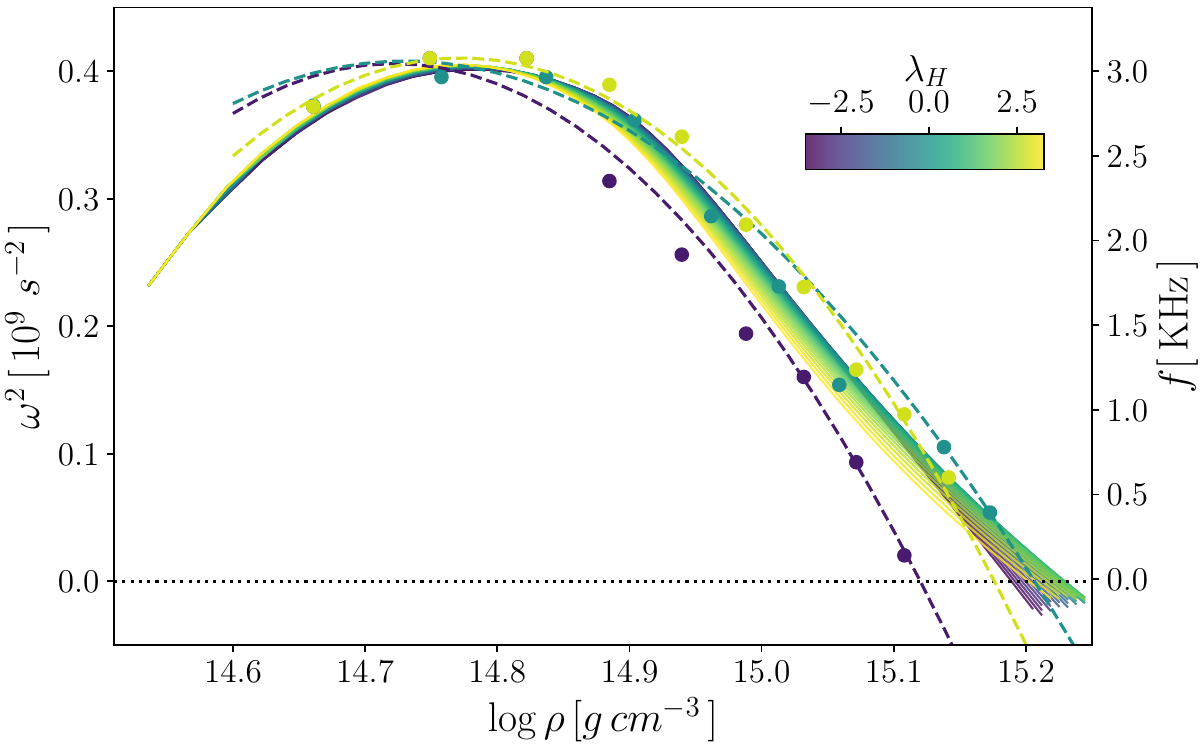}
    \includegraphics[width=0.95\linewidth]{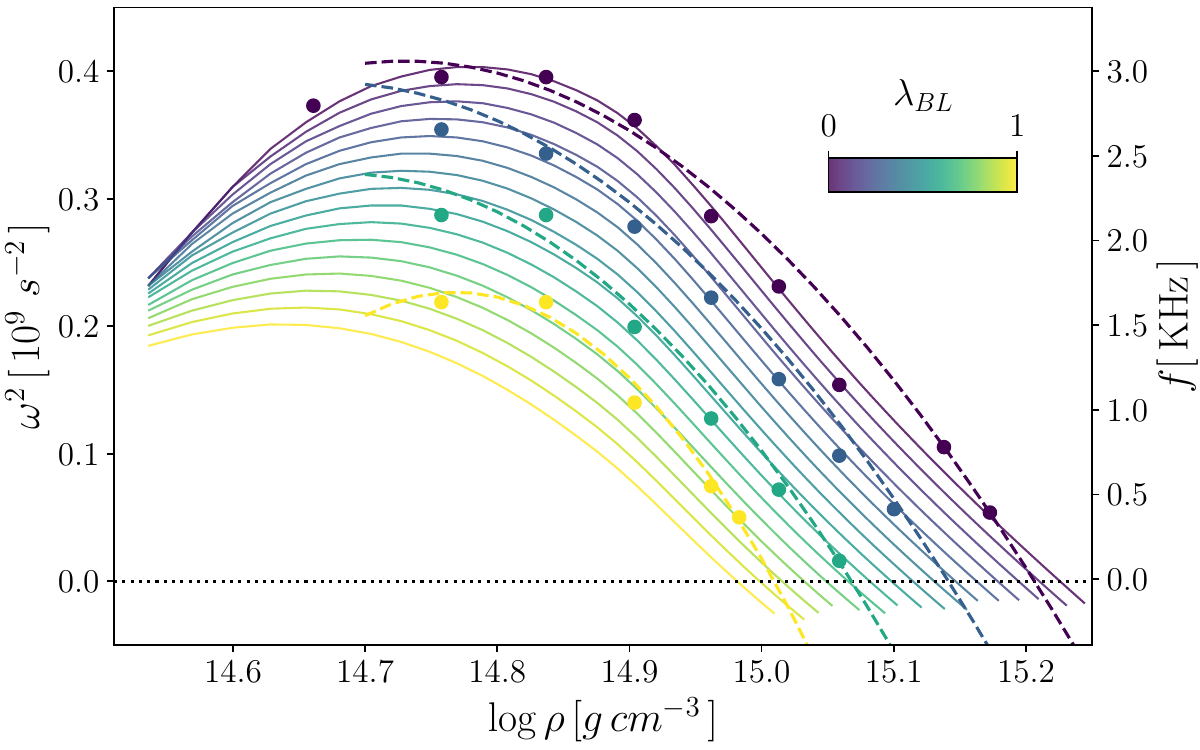}
    \includegraphics[width=0.95\linewidth]{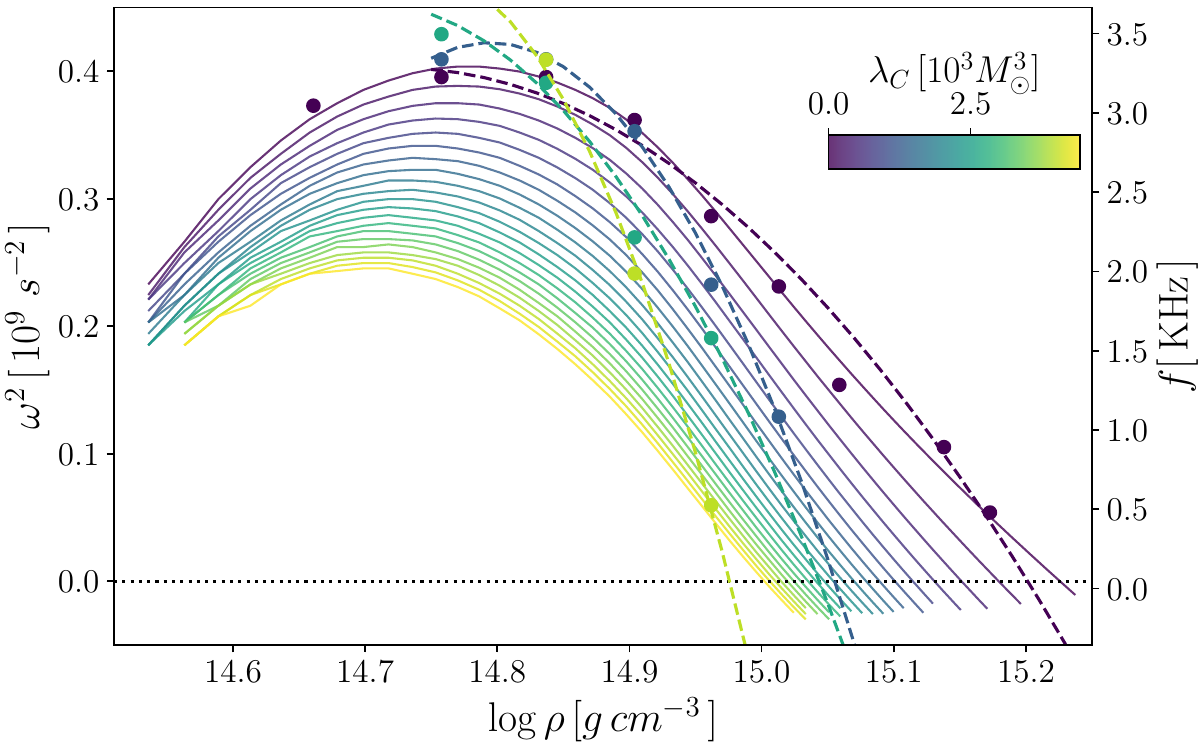}
    \caption{Squared frequency of the fundamental mode as a function of the star’s central rest-mass density for the Horvat (top panel), Bowers–Liang (middle panel), and Covariant (bottom panel) anisotropy models, using the QHC21 EOS. Colored dots represent results from non-linear simulations, while solid colored lines correspond to the linear stability analysis. \textcolor{black}{The dashed lines correspond to a fourth-order polynomial fit to the colored data points.} The color scale indicates the value of the anisotropy parameter in each model.}
    \label{fig:Omegaani}
\end{figure}

Given an anisotropy model, we generate a set of initial configurations by constructing equilibrium spherical models over a range of central rest-mass densities and anisotropy parameter values. After introducing a perturbation to the rest-mass density, we evolve each stellar model for approximately $10~\tau_{\rm sound} (\equiv R_{\rm star}/ c_{s,c})$, tracking the time evolution of both the fluid variables and the metric functions. For stable configurations, these quantities exhibit oscillations around their equilibrium values. In this regime, we aim to identify the fundamental mode of spherical radial oscillations. To extract its frequency, we compute the power spectral density (PSD) of the central rest-mass density as a function of time. The frequency of the fundamental mode is identified as the first prominent peak in the PSD.

For the three anisotropic models, Figure~\ref{fig:Omegaani} shows the squared frequency of the fundamental mode as a function of the star’s central rest-mass density, using the QHC21 EOS, as obtained from the non-linear numerical simulations, \textcolor{black}{$\omega_{\rm full}$} (colored dots). For comparison, the solid lines of Figure~\ref{fig:Omegaani} correspond to the squared frequency of the fundamental mode computed from the linear stability analysis of radial perturbation, \textcolor{black}{$\omega_{\rm lineal}$}, following the approach of \cite{2001MNRAS.322..389Y,2020EPJC...80..726P} (see also Appendix \ref{app:II}). 

\textcolor{black}{The impact of anisotropy on the fundamental-mode frequency depends on the functional form of the anisotropy parameter. In the Horvat model, the squared frequency is nearly independent of anisotropy at low central densities, but at higher densities it decreases as the difference between the tangential and radial pressures grows, regardless of whether the anisotropy is positive or negative\footnote{\textcolor{black}{Typically, for the Horvat model, the anisotropy is taken to be directly proportional to the compactness ($\sigma \propto 2m/r$). In this case, the expected behavior is recovered: the squared frequency of the fundamental mode is inversely proportional to the anisotropy constant at larger central densities (see, e.g., \citet{2020EPJC...80..726P}).}}. In the Bowers–Liang model, the squared fundamental frequency decreases with increasing anisotropy at fixed central density. The linear analysis for the covariant model predicts a behavior similar to that of the Bowers–Liang case; however, the numerical simulations show that the anisotropy increases the squared frequency at low central densities but causes a substantial reduction at higher densities. }

\textcolor{black}{Table~\ref{tab:sigmaO} lists the mean value of the relative error between the two appoches, computed as: $(\omega_{\rm lineal}^2-\omega^2_{\rm full})/\omega^2_{\rm lineal}$ , for each EOS and value  of the anisotropy parameter.} The linear and non-linear results agree well for the Horvat and Bowers–Liang anisotropy models, \textcolor{black}{particularly at low anisotropy}. For the covariant model, however, the linear formalism slightly underestimates the fundamental mode frequency. \textcolor{black}{From these results, we conclude that the Bowers–Liang model is the best captured by the linear approximation.}

\begin{figure*}
    \centering
    \includegraphics[width=0.32\linewidth]{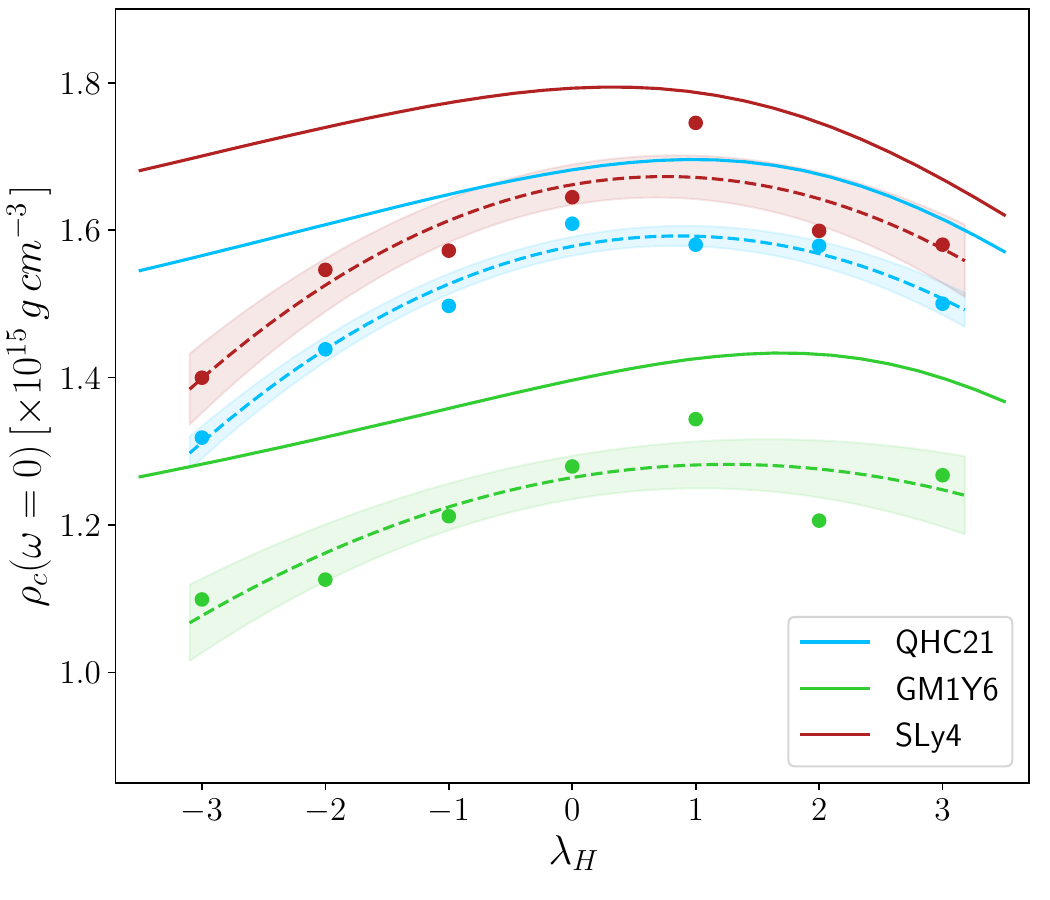}
    \includegraphics[width=0.32\linewidth]{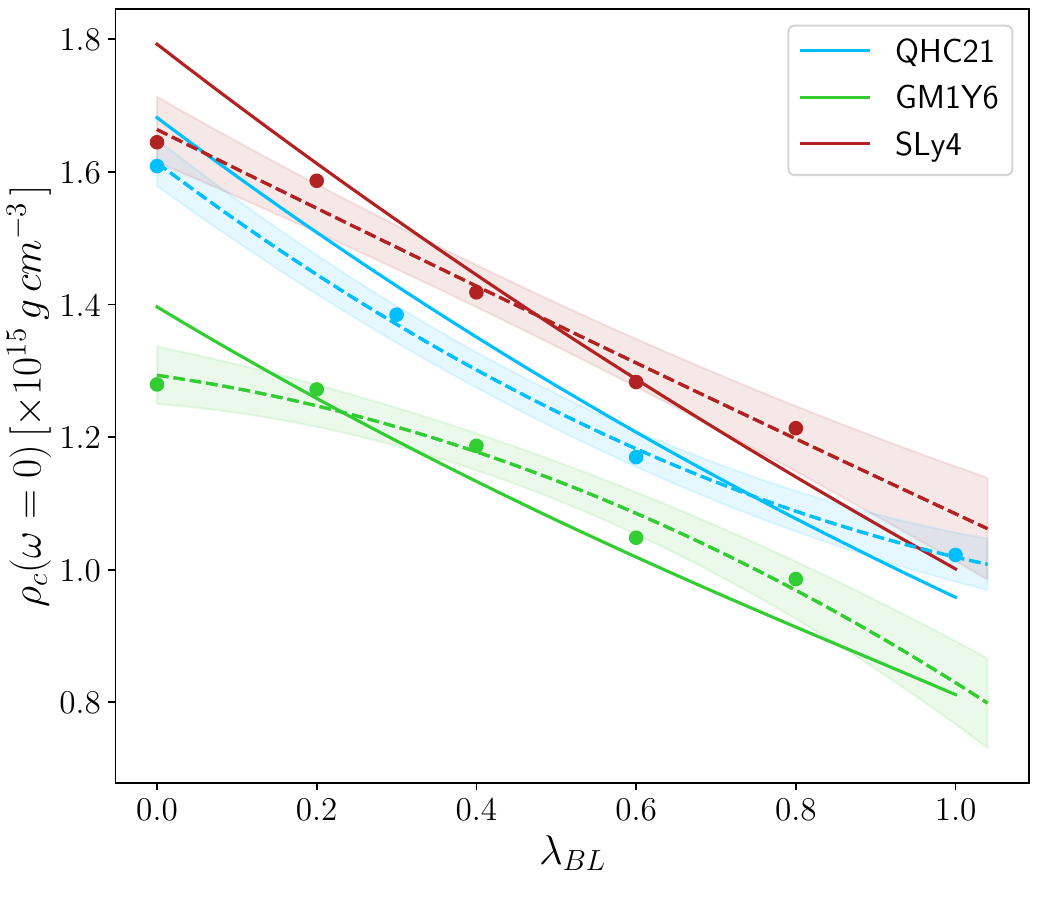}
    \includegraphics[width=0.32\linewidth]{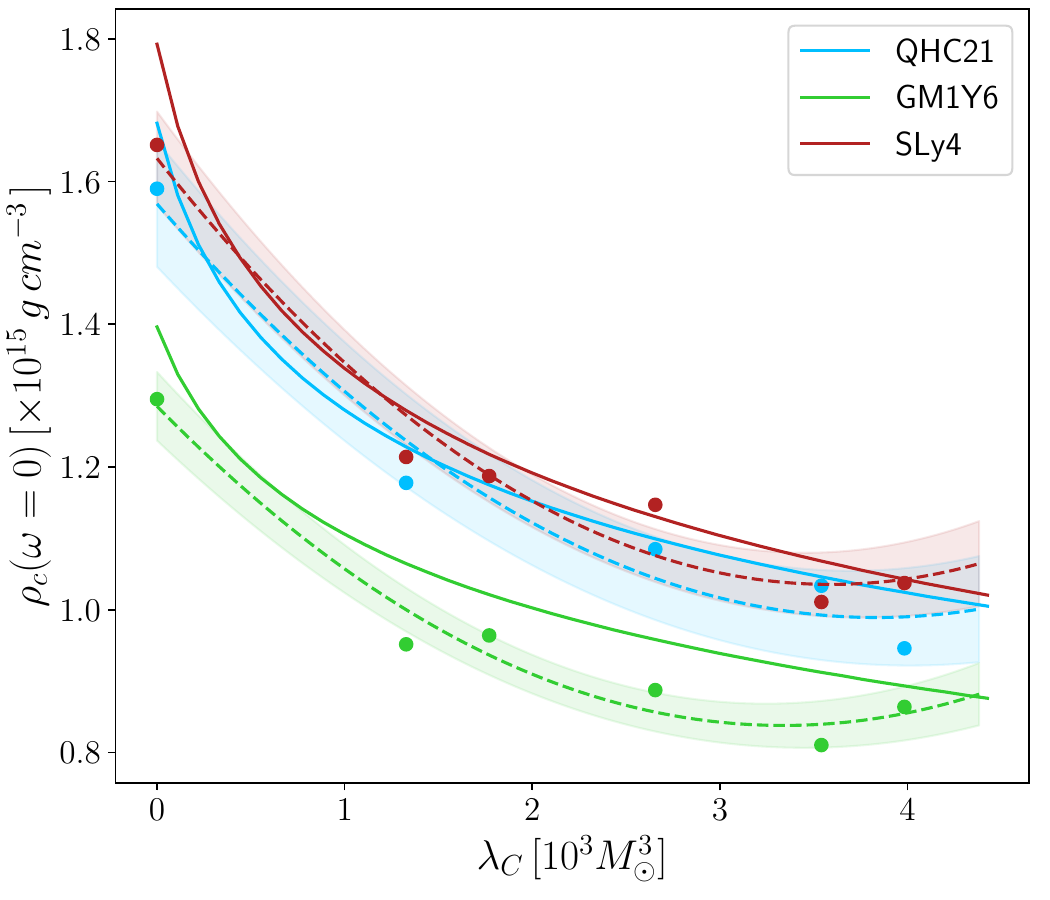}
    \caption{\textcolor{black}{Critical density as a function of the anisotropy parameter at which the fundamental mode frequency vanishes, shown for the Horvat (left), Bowers–Liang (center), and covariant (right) anisotropy models. Solid lines correspond to the semi-analytical results from the linear formalism. Dots are obtained from the numerical simulations, while the dashed line represents the analytical fit given by equation~(\ref{eq:rhoc_fit}). The shaded regions indicate the corresponding uncertainties.}}
    \label{fig:rhomax}
\end{figure*}

A  rigorous stability analysis involves studying the frequency of the fundamental mode of radial oscillations. For stable configurations, the squared frequency of the fundamental mode is positive ($\omega^2 > 0$), while for unstable ones it becomes negative ($\omega^2 < 0$). \textcolor{black}{As shown in Figure~\ref{fig:Omegaani},  for a fixed value of the anisotropy parameter, the fundamental frequency  increases with central density, reaches a maximum, and then decreases, eventually vanishing at a critical density that marks the onset of dynamical instability  ($\omega^2 = 0$). Near this point, however, the numerical simulations become noisy and unreliable. Then, to determine the critical density, we fit the numerical data for $\omega^2(\rho_c)$ with a fourth-order polynomial (dashed lines in Figure~\ref{fig:Omegaani}) and extract the density at which the fitted curve crosses zero \cite[see, e.g.,][for a similar procedure]{2011MNRAS.416L...1T}. The resulting values, shown as dots in Figure~\ref{fig:rhomax} for each EOS and anisotropy model, are compared with those obtained from the linear formalism. We find that the linear analysis generally overestimates the critical density, particularly for the Horvat model.  Overall, we see that the critical density decreases with increasing anisotropy in the Bowers–Liang and Covariant models, while in the Horvat case it decreases for positive anisotropy but increases as the anisotropy becomes more negative. For the Horvat model in particular, the value of the anisotropy parameter at which the critical density reaches its maximum depends on the chosen EOS.}

\textcolor{black}{
The neutral-stability line can be fitted with the following relation (dashed lines in Figure~\ref{fig:rhomax}):}
\begin{equation}\label{eq:rhoc_fit}
\frac{\rho_c(\omega=0, \lambda)}{10^{15} \, {\rm g\, cm}^{-3}}= a+ b\lambda + c\lambda^2
\end{equation}

\textcolor{black}{where the coefficients $a$, $b$ and $c$ depend on the anisotropy model and the EOS, and are listed in Table~\ref{tab:rhoc_coeff} (see Appendix~\ref{app:I} for the convergence of the neutral-stability line)
}


\begin{table*}
    \centering    
    \begin{tabular}{K{0.8cm}|ccc||K{0.8cm}|ccc||c|ccc }
    \hline
    \multicolumn{4} {c||} {{\bf Hovart}} &\multicolumn{4} {c||} {{\bf Bowers-Liang}}& \multicolumn{4} {c} {{\bf Covariant}} \\
      \hline
      &   & &  &  &   &  &  &   &   &  &\\
      $\lambda_H$ & $\,$ QHC21  &$\,$  GM1Y6  & $\,$ SLy4 $\,$ & $\lambda_{BL}$ &$\,$  QHC21  & $\,$ GM1Y6 &$\,$  SLy4 & $\lambda_{C}$ & $\,$ QHC21  & $\,$ GM1Y6  &$\,$  SLy4 $\,$\\ 
      &   & &  &  &   &  &  & $[10^3 M_\odot^3]$  &   &  &\\
      \hline
       &   & &  &  &   &  &  &   &   &  &\\
        $-3$ & $0.111$  & $0.121$ & $0.128$ & $0.2$ &  - & $0.018$ & $0.073$ & $1.3$ &  $0.168$  & $0.029$ & $0.049$\\
        $-2$ & $0.041$  & $0.099$ & $0.046$ & $0.3$ &  $0.001$ & -  & - & $1.7$ &  -  &$0.038$  &$0.120$ \\
        $-1$ & $0.038$  & $0.050$ & $0.025$  & $0.4$ & -  & $0.009$ & $0.003$ & $2.6$ & $0.453$  & $0.115$ &$0.317$\\
        $0$  & $0.005$  & $0.001$  & $0.002$  & $0.6$ & $0.012$  & $0.016$ & $0.004$& $3.5$ & $0.527$  &$0.216$  &$0.535$\\
        $2$  & $0.025$  & $0.115$ & $0.029$   & $0.8$ &   &  $0.047$ & $0.022$ & $3.9$ & $0.828$   & $0.264$ &$0.646$\\
        $1$  & $0.050$  & $0.113$ & $0.009$  & $1.0$ & $0.378$  &-  &- &  &    &  &\\
        $3$  & $0.047$  & $0.026$ & $0.161$  &      &   &  &  &      &   &  &\\
         &   & &  &  &   &  &  &   &   &  &\\
        \hline \hline   
    \end{tabular}
    \caption{\textcolor{black}{Relative error between the fundamental frequency obtained from the linear stability analysis and that from the full GR simulation}}
    \label{tab:sigmaO}
\end{table*}

\begin{figure*}
    \centering
    \includegraphics[width=0.95\linewidth]{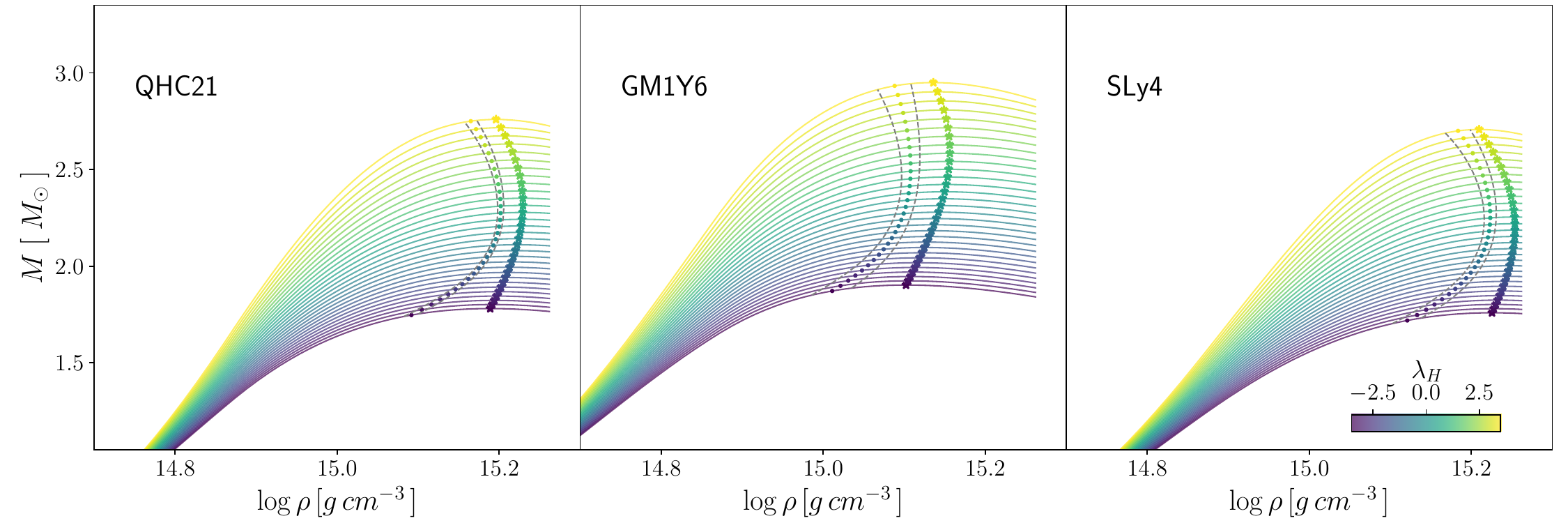}
    
    \includegraphics[width=0.95\linewidth]{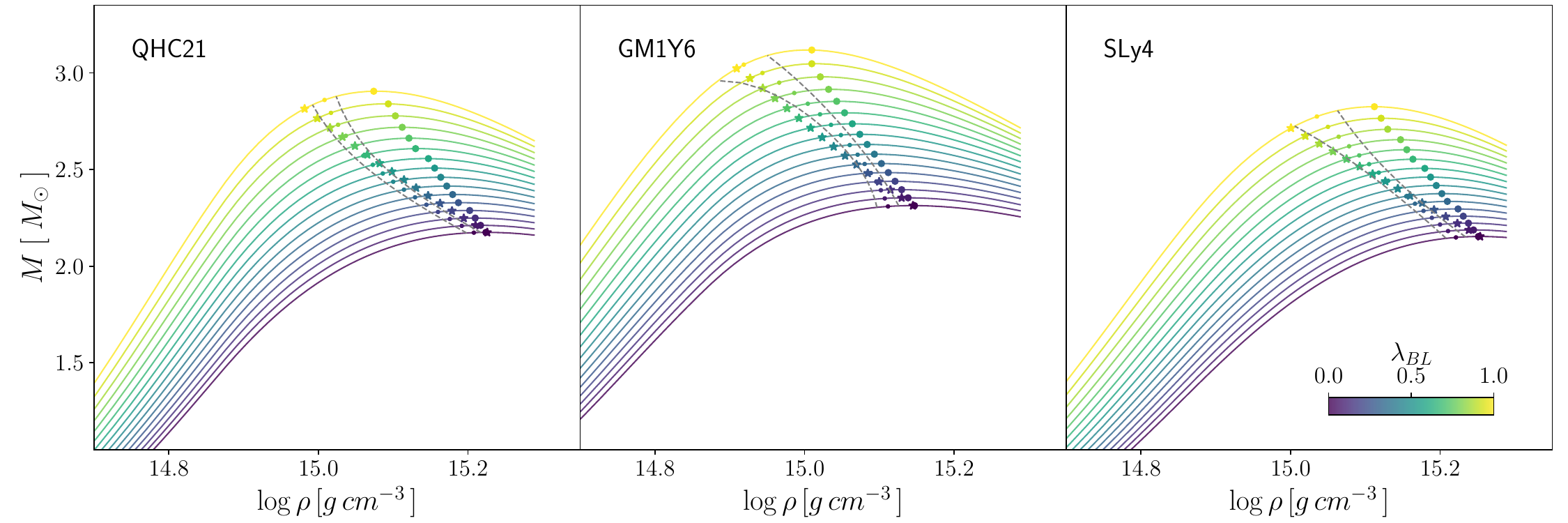}

    \includegraphics[width=0.95\linewidth]{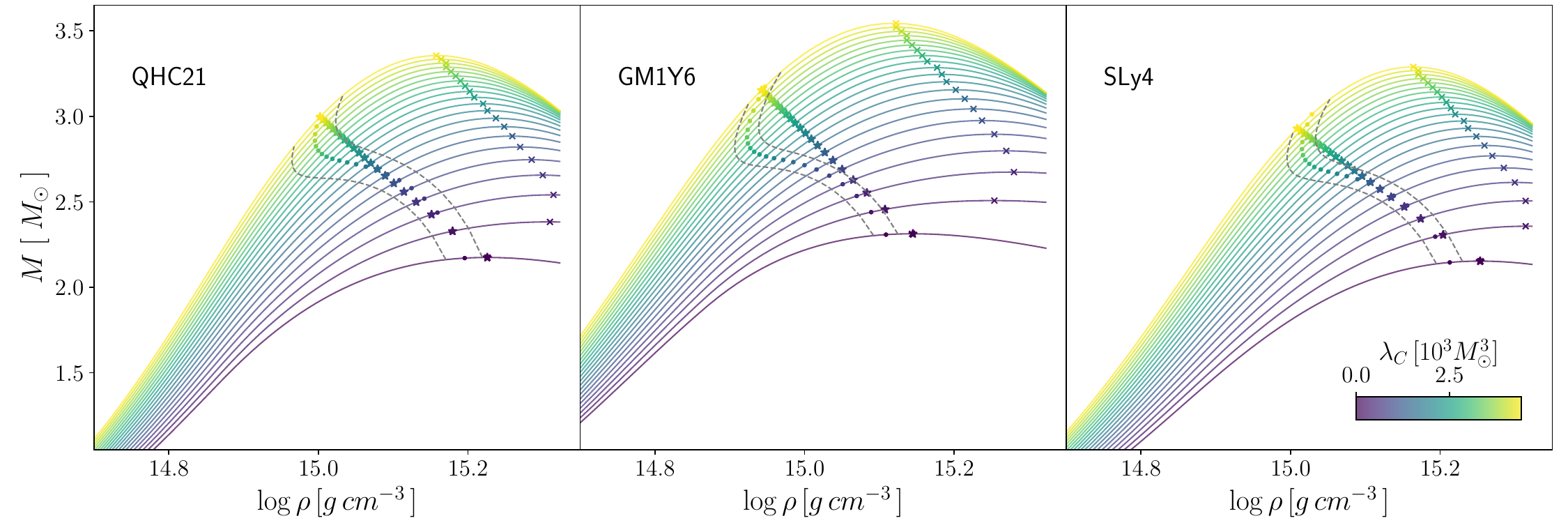}
    \caption{\textcolor{black}{Gravitational mass} of anisotropic spherical configurations as a function of the central rest-mass density, for the three anisotropy models: Horvart  (upper panel), Bowers-Liang (middle panel ), and Covariant model (bottom panel); and three NS EOS: QHC21, GM1Y6, and SLy4 EOS.  Star-shaped  points to the configuration with vanishing frequency of the fundamental mode as computed from the linear stability analisys, \textcolor{black}{circular markers corresponds to the analytical fit given in equation~(\ref{eq:rhoc_fit}) and  obtained from the numerical simulations}, and cross markers denote the point at which the star reaches its maximum mass. \textcolor{black}{Dash gray lines enclose the corresponding uncertainties of the analytical fit}. The color scale corresponds to the value of the anisotropic parameter.}
    \label{fig:Cani}
\end{figure*}

\begin{table}[h!]
    \centering
    \begin{tabular}{c|K{1.2cm}|cccc}
    \hline
      Model & EOS   & $a$ & $b$& $c$ &$\chi^2$\\ \hline
        &   &  &  &  & \\
     Hovart    & QHC21   & $\,1.578$ & $\,0.032$  & $\,-0.019 $ & $0.0003$ \\
         & GM1Y6   & $\, 1.264$ & $\, 0.028$ & $\, -0.011$ &  $0.0016$\\
         & SLy4   & $\, 1.661$  & $ 0.029$ &$-0.019$ & $ 0.0013$ \\ \hline
         &   &  &  &  & \\
     Bowers-Liang   & QHC21   & $1.613$ & $-0.905$ & $0.310$6 & $0.0001$\\
         & GM1Y6  & $\, 1.293$ & $\, -0.172$ & $\, -0.291$ & $0.0005$ \\ 
         & SLy4  & $1.663$ & $-0.597$ &$0.017$ & $0.0006$ \\  \hline
         &   &  &  &  & \\
    Covariant    & QHC21   & $1.568$ &$-0.302$  &$0.039$ &$0.0018$ \\
         & GM1Y6   & $1.285$ & $-0.268$ & $0.040$ & $0.0008$\\
         & SLy4   & $1.632$ & $-0.332$ & $0.046$ & $0.0015$ \\ \hline\hline
    \end{tabular}
    \caption{ \textcolor{black}{Fit parameter of equation~(\ref{eq:rhoc_fit}).  that in the Horvat and Bowers–Liang models, the parameters $a$, $b$ and $c$ are dimensionless, whereas in the Covariant model, $a$ is dimensionless and $b$ and $c$ have dimensions of $[10^3 M_\odot^3]$ and $[(10^3 M_\odot^3)^2]$, respectively.}}
    \label{tab:rhoc_coeff}
\end{table}

A commonly used, though simplified, criterion for determining stability against gravitational collapse is based on the behavior of the stellar mass as a function of central density. Configurations satisfying $dM/d\rho_c > 0$ are typically regarded as stable, while those with $dM/d\rho_c < 0$ are classified as unstable. The central density at which the mass reaches its first maximum is often used to define a critical threshold for stability. However, this condition is necessary but not sufficient for a complete stability assessment, as illustrated in Figure~\ref{fig:Cani}. The figure shows the \textcolor{black}{gravitational mass} of anisotropic spherical configurations as a function of the central rest-mass density, for the three anisotropy models and the three EOSs considered in this work.  The figure clearly illustrates that introducing anisotropy allows configurations to reach higher \textcolor{black}{mass} values than in the isotropic case. Star-shaped markers indicate the central density at which the fundamental mode frequency vanishes \textcolor{black}{obtained from the linear stability analysis}, while cross markers denote the point where the stellar mass reaches its maximum. In the isotropic case, as well as for the Horvat anisotropy model, both stability criteria coincide. However, for the Bowers–Liang and Covariant models, the onset of dynamical instability occurs at lower central densities than the maximum mass point, highlighting the importance of performing a full dynamical analysis when assessing the stability of anisotropic stars\footnote{It is worth noting that other studies, such as \citet{2020EPJC...80..726P}, report that for the Covariant model both criteria also coincide; however, their analysis uses $f(\epsilon) = \epsilon$ in equation~(\ref{eq:sigma_C}).}.  

\textcolor{black}{Figure~\ref{fig:Cani} also shows the fit to the neutral-stability line derived from the numerical simulations (see equation~\ref{eq:rhoc_fit}), with the corresponding uncertainties indicated by the dashed gray lines. For the Bowers–Liang model, the linear prediction lies within this uncertainty band, indicating that the linear analysis provides a good approximation to the neutral-stability line. In contrast, for the Horvat and Covariant models, the linear stability analysis overestimates the critical density as the anisotropy parameter increases. From these results, we find that the maximum mass of a configuration that is stable against radial perturbations is approximately $0.05\,M_\odot$ smaller than the turning-point prediction for the Horvat model, less than $0.1\,M_\odot$ smaller for the Bowers–Liang model, and up to $0.4\,M_\odot$ smaller for the Covariant model}

In general, anisotropy can increase the maximum mass of a \textcolor{black}{stable NS} by up to $\sim 30\%$ compared to the isotropic case.

\subsection{\label{sec:collapse} Gravitational Collapse}

\begin{figure}[t!]
    \centering
    \includegraphics[width=0.99\linewidth]{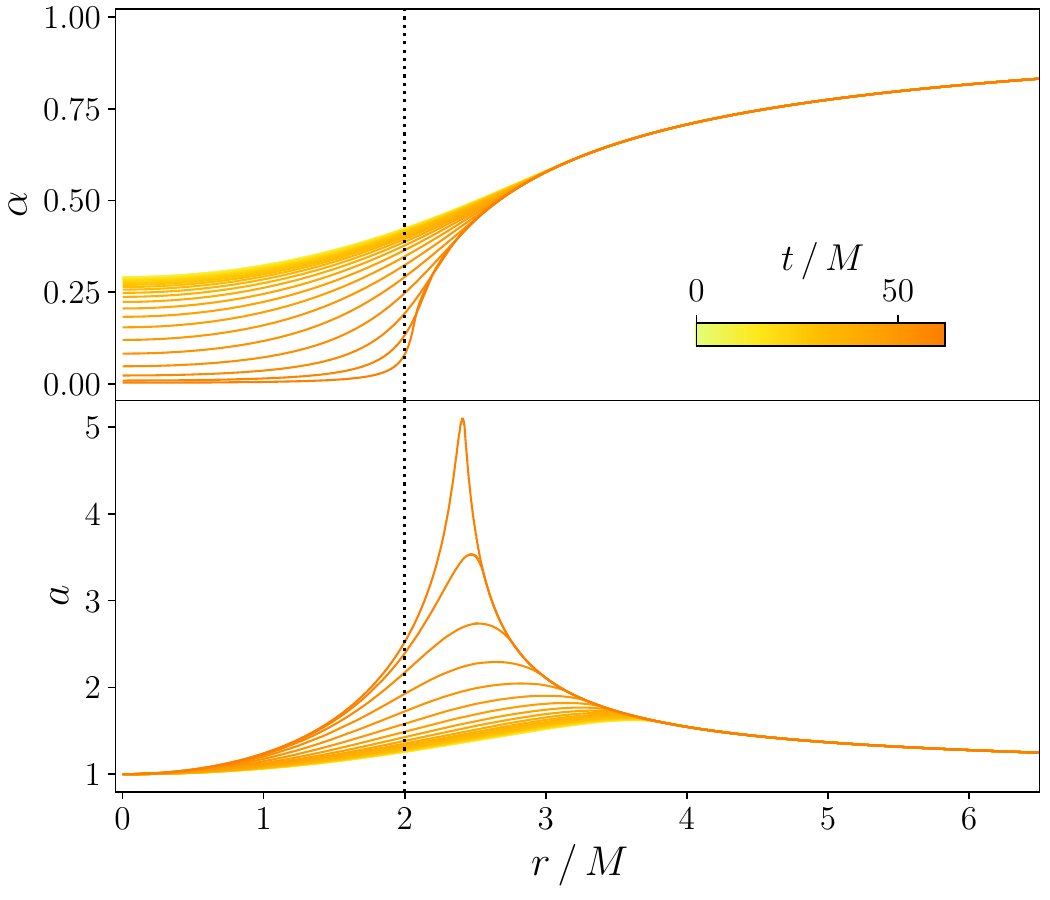}
    \caption{ Radial profiles of the metric functions $\alpha$ (top panel) and $a$ (bottom panel) for an $2.17~M_\odot$ unstable isotropic configuration using the QHC21 EOS. The color scale corresponds to different time steps.}
    \label{fig:gg_iso}
\end{figure}

\begin{figure}
    \centering
    \includegraphics[width=0.95\linewidth]{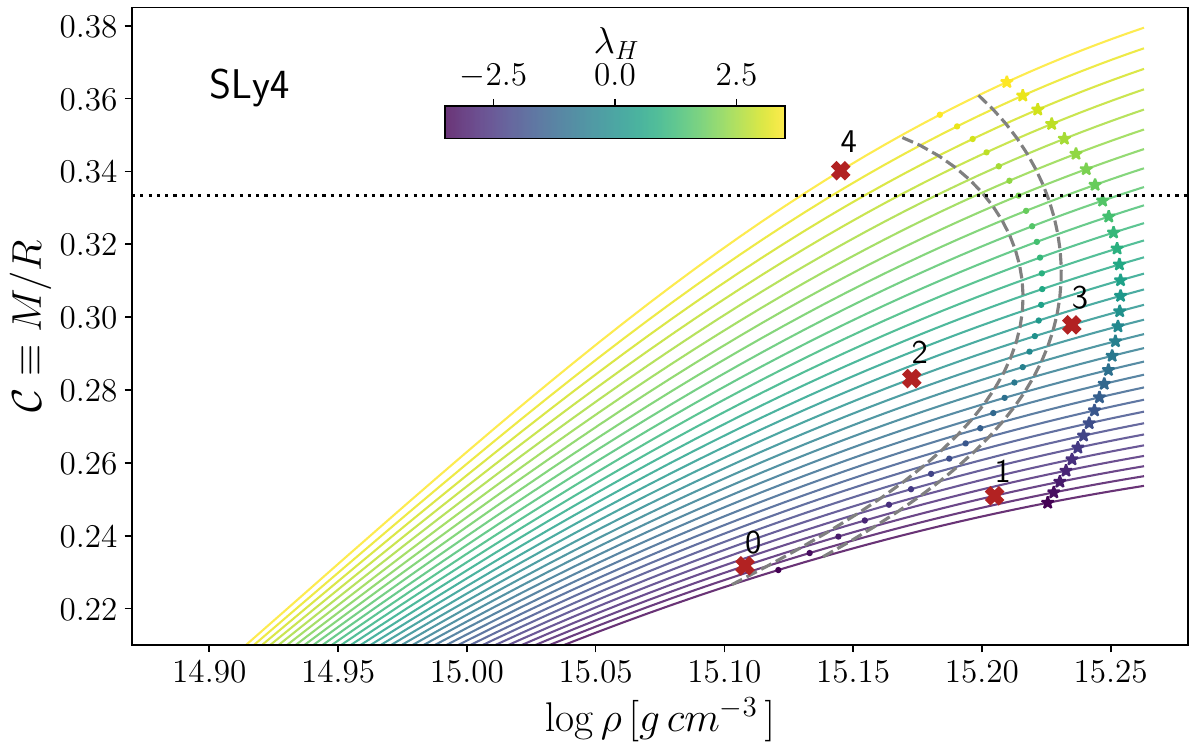}
    \includegraphics[width=0.95\linewidth]{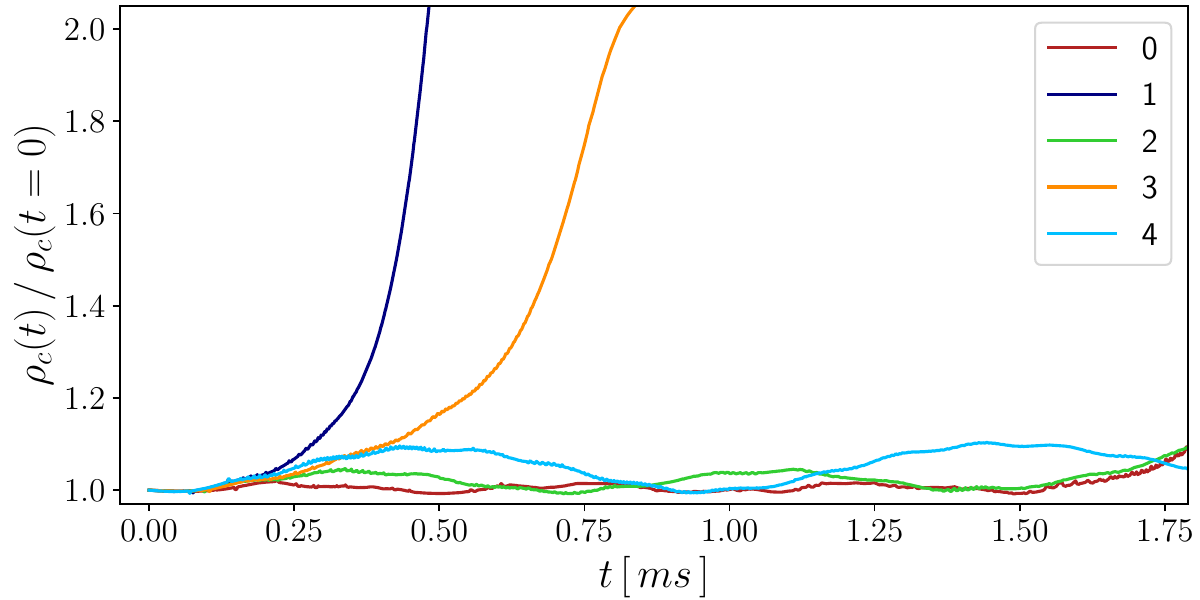}
    \caption{ \textcolor{black}{Upper: Compactness of anisotropic spherical configurations as function of the central rest-mass density for the Horvart anisotropy model and the SLy4 EOS. Bottom: Central density time evolution} }
    \label{fig:comp_h}
\end{figure}

\begin{figure}[t!]
    \centering
    \includegraphics[width=0.95\linewidth]{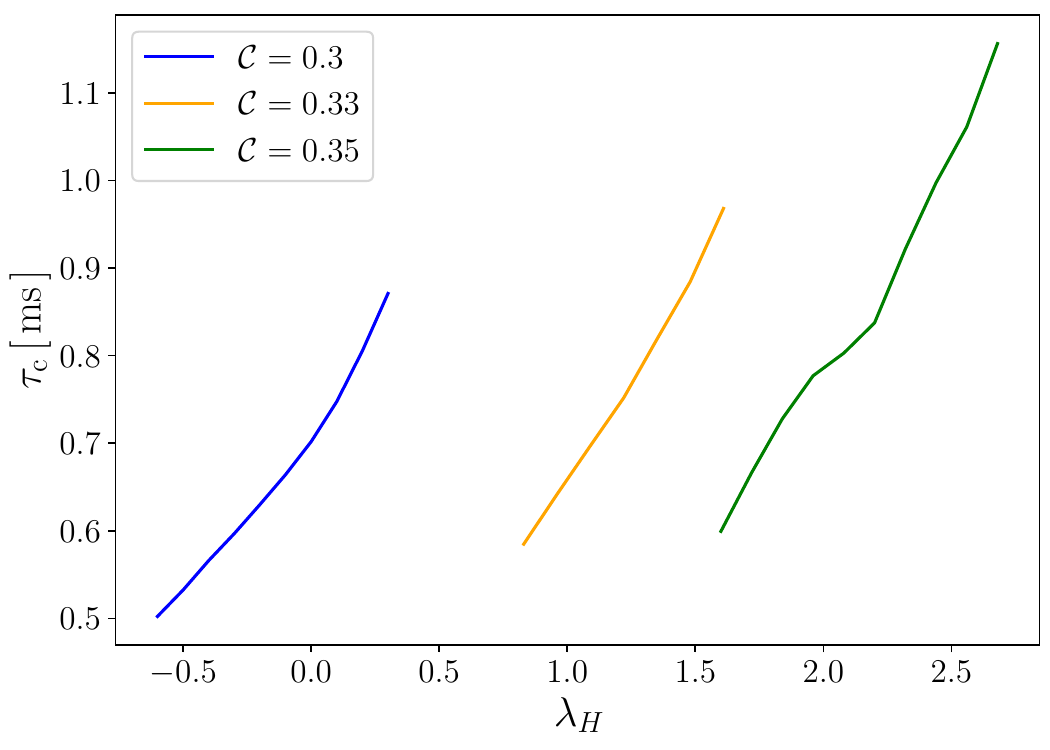}
    \caption{Estimated collapse time of unstable anisotropic configurations decribed by the Horvart model  and using the QHC21 EOS.}
    \label{fig:TimeC}
\end{figure}

If the initial configuration is unstable, a perturbation triggers gravitational collapse, ultimately leading to the formation of a BH. Figure~\ref{fig:gg_iso} shows the radial profiles of the metric functions, $\alpha=\sqrt{-g_{tt}}$ and $a=\sqrt{g_{rr}}$, at different times for an unstable isotropic configuration with an initial gravitational mass of $2.17~M_\odot$, using the QHC21 EOS.

As the collapse progresses, the radial metric function $\alpha$ develops a sharp peak that cannot be adequately resolved, resulting in a rapid increase in numerical error. As a result, the simulation eventually crashes, preventing us from capturing the final stages of BH formation. Nonetheless, the evolution exhibits characteristic signatures of gravitational collapse and therefore formation of an apparent horizon: the function $\alpha$ collapses to zero, while the function $a$ diverges near the location of the horizon.

\textcolor{black}{To verify the accuracy of the neutral-stability line fit given in equation~(\ref{eq:rhoc_fit}), we examined the time evolution of the configurations marked in the top panel of Figure~\ref{fig:comp_h} for the Horvat anisotropy model and the SLy4 EOS. As shown in the bottom panel, configurations lying to the left of the fitted neutral-stability line remain stable under radial perturbations, while those to the right become unstable. Notably, configurations “one’’ and “three’’ are predicted to be stable by the turning-point criterion, yet the dynamical simulations reveal that they are in fact unstable.}

\textcolor{black}{The top panel of Figure~\ref{fig:comp_h} also shows a horizontal black line corresponding to compactness $C = 1/3$. Configurations above this line have compactness values exceeding $1/3$, implying the existence of a photon ring outside the star. Such extreme compactness can be reached only in the presence of anisotropy (and will be examined further in Section~\ref{sec:ecos}). However, for the Bowers–Liang and covariant anisotropy models, all configurations with $C > 1/3$ are unstable against radial perturbations. Only in the Horvat model is it possible to construct stable configurations with compactness greater than $1/3$, such as configuration “four’’.}


Finally, to illustrate the effect of anisotropy on the collapse of unstable configurations, Figure~\ref{fig:TimeC} presents an estimate of the collapse time, $\tau_c$, for unstable anisotropic configurations constructed using the Horvat anisotropy model and the QHC21 EOS. Increasing the anisotropy parameter results in configurations with larger masses, which in turn exhibit longer collapse times. Here, $\tau_c$ is defined as the elapsed time from the start of the simulation to the point at which the numerical evolution can no longer proceed due to numerical errors. For consistency, we take $\tau_c$ to be $90\%$ of the time required for the apparent horizon to form. A systematic study of apparent horizon formation after stellar collapse and the influence of anisotropy requires a change of variables to overcome numerical limitations and extend the evolution. This is beyond the present work's scope and will be addressed in future studies.

\section{\label{sec:ecos} Gravitational wave echoes}


\begin{figure}
    \centering
    \includegraphics[width=0.95\linewidth]{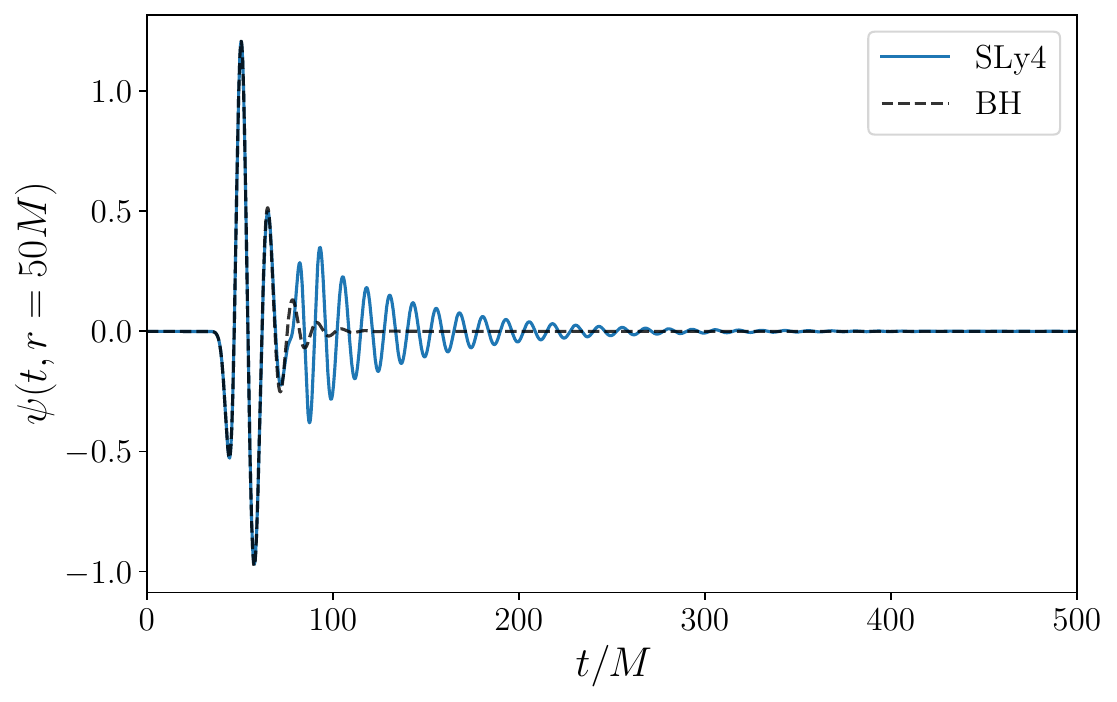}
    \includegraphics[width=0.95\linewidth]{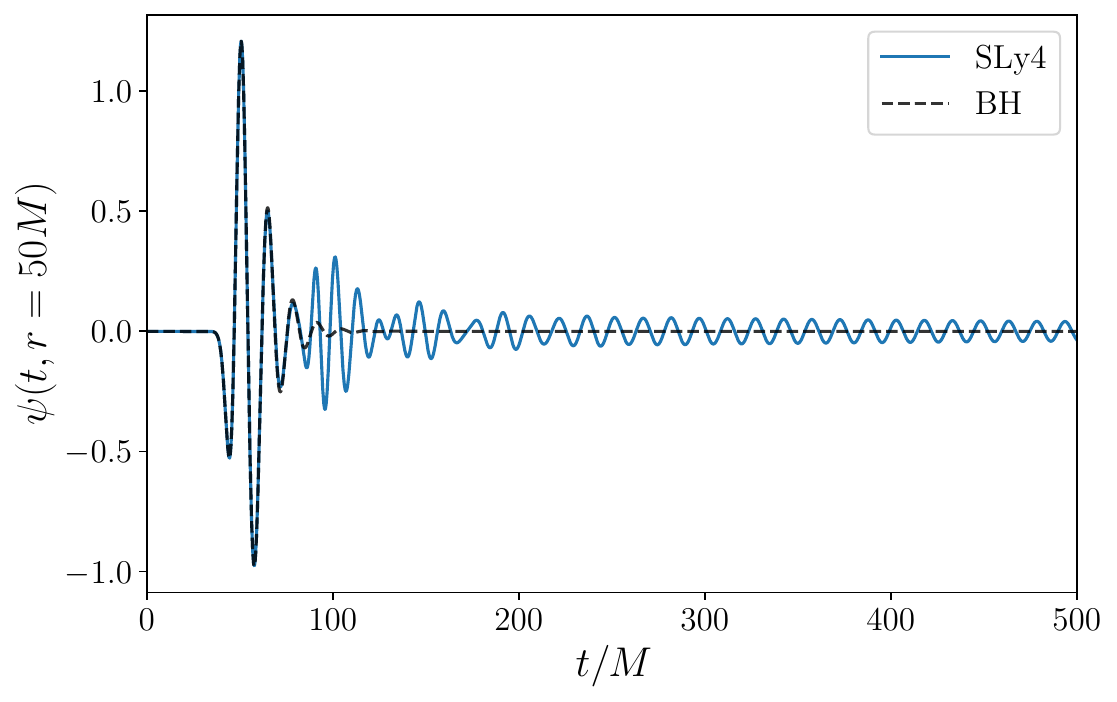}
    \includegraphics[width=0.95\linewidth]{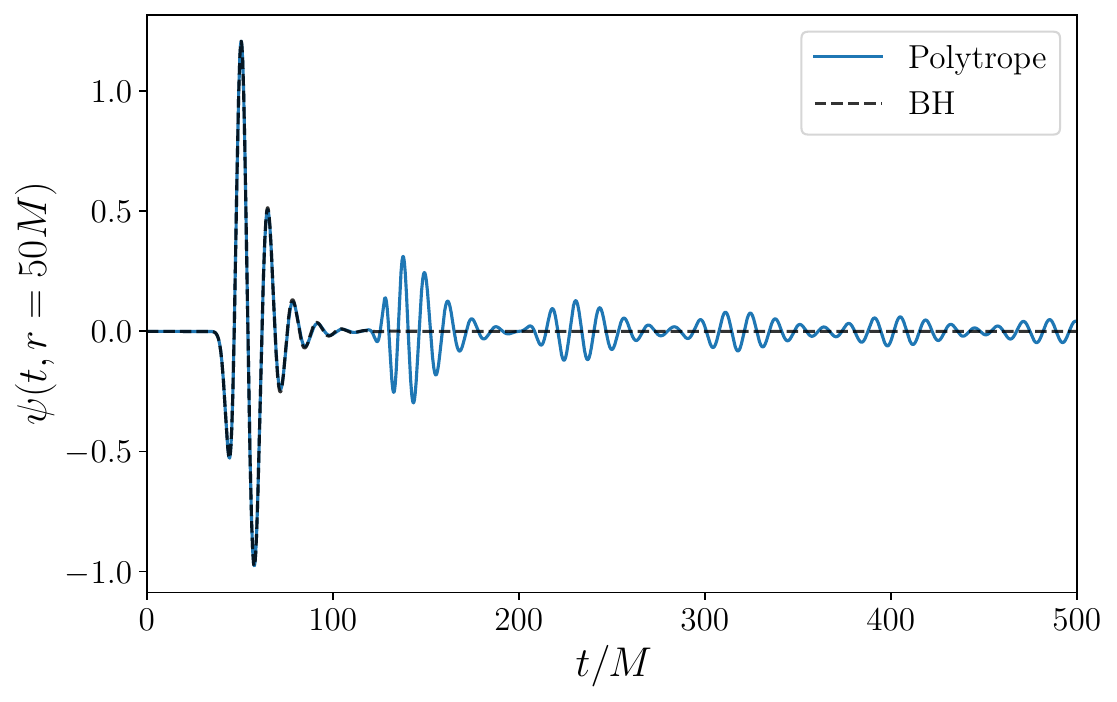}
    \caption{
    Gravitational-wave signals from anisotropic NSs with different properties. Top: A physical and stable configuration with EOS SLy4, compactness $\mathcal{C}=0.34$, and Horvat anisotropy $\lambda_H = 2.5$. Middle: An unstable physical configuration with EOS SLy4, compactness $\mathcal{C}=0.409$, and $\lambda_H = 7.5$, which shows no echoes despite the high compactness. Bottom: A non-physical configuration modeled with a polytropic EOS ($\Gamma = 3.2$, $K=100$), compactness $\mathcal{C}=0.4$, and Horvat anisotropy $\lambda_H = 0.24$, which exhibits GW echoes.}
    \label{fig:psiQHC}
\end{figure}
\begin{figure}
    \centering
    \includegraphics[width=0.9\linewidth]{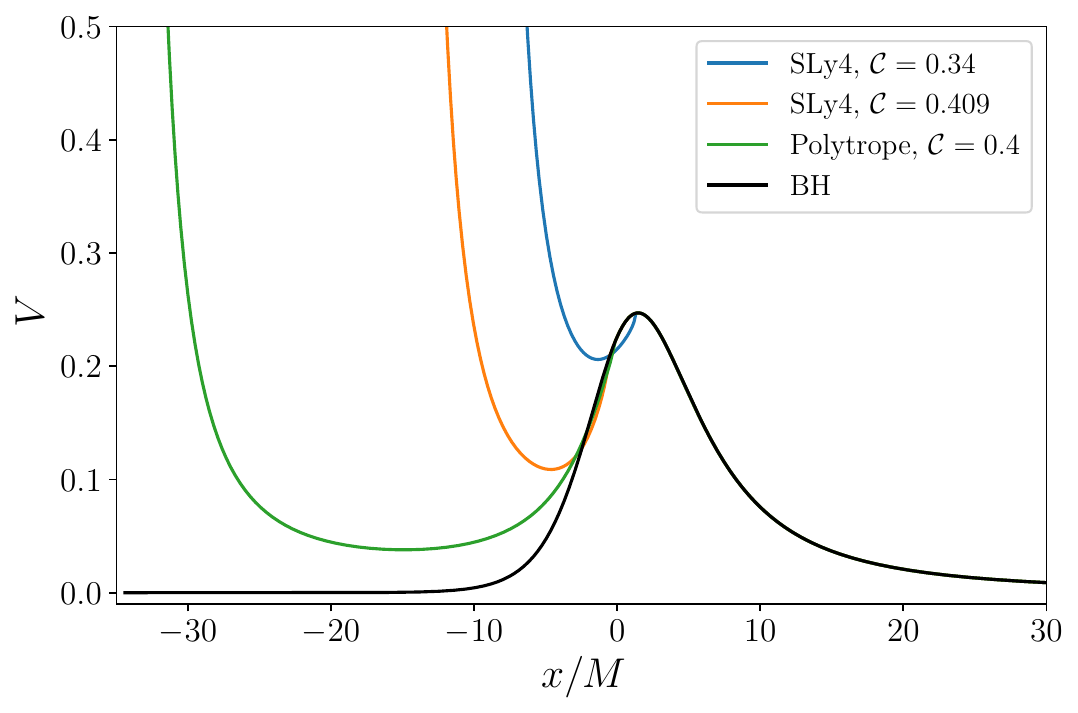}
    \caption{Comparison of the potential \eqref{eqn:Veff} for selected compact stars following the anisotropy model of Horvat. The waveform is shown in Figures \ref{fig:psiQHC}.
    }
    \label{fig:Veff}
\end{figure}

According to \cite{Cardoso:2016rao}, highly compact objects can produce a BH-like GW ringdown. The full waveform, however, may also include a sequence of secondary pulses called echoes. A ringdown without echoes would support the formation of an event horizon, while detecting echoes could indicate a horizonless, exotic compact object.

As noted in \cite{Price:2017cjr, Mark:2017dnq}, echoes are not a universal feature of compact objects with compactness $\mathcal{C} > 1/3$. This raises the question of whether echoes are a generic signature of GWs from realistic anisotropic NSs. If true, the detection of echoes would give hints towards the existence of anisotropic NSs. We will show that although some of the physically motivated NS configurations discussed above satisfy $\mathcal{C} > 1/3$, they do not produce echoes.

To show this, we followed the standard approach of modeling a massless scalar field $\Psi$ on a static NS background as a proxy for the GW emission from a truly perturbed anisotropic NS. Specifically, we solved the Klein-Gordon equation $\nabla_\alpha\nabla^\alpha \Psi = 0$ in a curved spacetime. Applying the usual spherical harmonic decomposition $\Psi = \sum Y_{\ell m}(\theta,\phi)\, \psi_{\ell m}(t,r)$ leads to the following wave equation:
\begin{equation}
    -\partial_{tt} \psi + \partial_{xx} \psi - V\,\psi = 0, \label{eqn:waveEQpsi}
\end{equation}
where $x$ is the tortoise coordinate defined by $dr/dx = \alpha/a$, and the potential $V$ is given by
\begin{equation}
    V = \alpha^2(r) \frac{\ell(\ell+1)}{r^2} + \frac{1}{2r}\frac{d}{dr}\left[\frac{\alpha^2(r)}{a^2(r)}\right], \label{eqn:Veff}
\end{equation}
with $\ell$ denoting the multipole number.

We numerically solved Eq.~\eqref{eqn:waveEQpsi} using a leapfrog integrator \cite{press2007numerical}, imposing boundary conditions that ensure regularity at the center of the star $x_c$, and outgoing radiation at spatial infinity:
\begin{align}
    \psi|_{x=x_c} &= 0, \\
    \left.\left(\partial_t \psi + \partial_x \psi\right)\right|_{x=\infty} &= 0,
\end{align}
We chose a Gaussian profile for the initial conditions, 
\begin{align}
    \psi|_{t=0} &=0,\\
    \partial_t \psi|_{t=0} &=  e^{-\frac{(x-x_0)^2}{2\sigma^2}}.
\end{align}

The waveform for the $\ell=2$ multipole extracted at $x=50M$ for representative cases is shown in Figure~\ref{fig:psiQHC}.
We found no echoes, even for the most compact, stable, and physically motivated configurations, those with subluminal radial and tangential sound speeds.

For configurations with $\mathcal{C} > 1/3$, the early part of the signal (the precursor) is universal, as it arises directly from the initial perturbation. The subsequent part of the waveform closely matches the quasinormal mode (QNM) ringdown of a Schwarzschild BH. The number of oscillation cycles in this phase depends on the separation between the potential peak $V$ (located at $r = 3M$) and the reflecting surface at the center of the star. For the physical configurations we studied, this reflecting surface lies very close to the potential peak, which causes the reflection to promptly interfere with the BH–like part of the signal, effectively suppressing any distinct echo structure; see Figure \ref{fig:Veff}. This feature can be seen in the top panel of Figure~\ref{fig:psiQHC}.

Next, we solved the wave equation for configurations beyond stability to achieve higher compactness values. We found that even for a configuration with $\mathcal{C}\approx 0.4$, the echoes are absent, as shown in the middle panel of Figure~\ref{fig:psiQHC}. Since in this case the reflection surface is farther from the potential peak, more cycles of the BH QNM-like signal are observed. However, the time delay is not sufficient for the formation of localized repeated signals. Instead, a long tail is formed from the superposition of the multiple reflections.

For comparison, we obtained the waveform for a polytrope NS, which is shown in the bottom panel of Figure~\ref{fig:psiQHC}. The waveform exhibits echoes similar to those found in \cite{2019PhRvD..99j4072R}. This is a consequence of the fact that in this case the potential is steeper and its reflecting surface is farther from the peak; see the green curve in Figure~\ref{fig:Veff}. However, both the radial and the tangential sound speeds are superluminal for this configuration. We conjecture that, for sufficiently compact configurations, the EOS significantly influences echo production. Echoes arise in configurations of equal compactness under a single polytropic EOS.
%


\section{\label{sec:discussion} Discussions and Conclusions}
We have constructed anisotropic NS configurations using three distinct ansatz to describe the pressure anisotropy: the Horvat, Bowers–Liang, and Covariant models, together with three EOS representing different particle compositions: nucleons, nucleons with hyperons, and quarks. Each EOS is described by a GPP parametrization with a continuous sound speed. The stability of these configurations is assessed through their dynamical evolution using a fully non-linear relativistic solver. For stable configurations, we have computed the oscillation spectrum and identified the fundamental mode frequency, which serves as a key indicator of dynamical stability. \textcolor{black}{We have presented fits for the the neutral-stability line as a function of the anisotropy parameter, given in equation~(\ref{eq:rhoc_fit}).}

Our results show that, in both the isotropic and Horvat models, the vanishing of the fundamental mode \textcolor{black}{in the linear formalism} coincides with the maximum-mass point, in agreement with the standard turning-point criterion for stability. In contrast, for the Bowers–Liang and Covariant models, instability arises at lower central densities. This indicates that the conventional turning-point method may not reliably predict the onset of dynamical instability when pressure anisotropy is present, remaining a necessary but not sufficient condition for stability.

Introducing pressure anisotropy generally increases the maximum NS mass given an EOS by up to $\sim 30\%$ relative to the isotropic case and allows for more compact stellar configurations. Moreover, when anisotropic stars do become unstable, they exhibit longer collapse times compared to isotropic counterparts of the same compactness, which could have observable implications in scenarios such as NS mergers or delayed BH formation.

Finally, we investigated the production of GW echoes as a possible observational signature of these compact NS. Our analysis, using a massless scalar field as a first approximation, reveals that no physically viable, stable, and causal configurations produce echoes. In particular, this holds even for highly compact objects that meet the condition $\mathcal{C}>1/3$. We conjecture that the formation of echoes depends critically on the star's internal structure, implying that high compactness is a necessary but not sufficient condition. Consequently, our results suggest that GW echoes may not be a reliable probe for the existence of these anisotropic compact NSs.
\appendix

\section{Code convergence}\label{app:I}

To assess the convergence of the code, we performed a resolution study by evolving a stable, isotropic star with mass $1.6\,M_\odot$ and the QHC21 EOS, varying the spatial resolutions of the grid. The simulations were carried out with radial spacings $\Delta r = \{0.0026, 0.0051, 0.0103, 0.0205\}\,M_\odot$.

\begin{figure}
    \centering
    \includegraphics[width=0.95\linewidth]{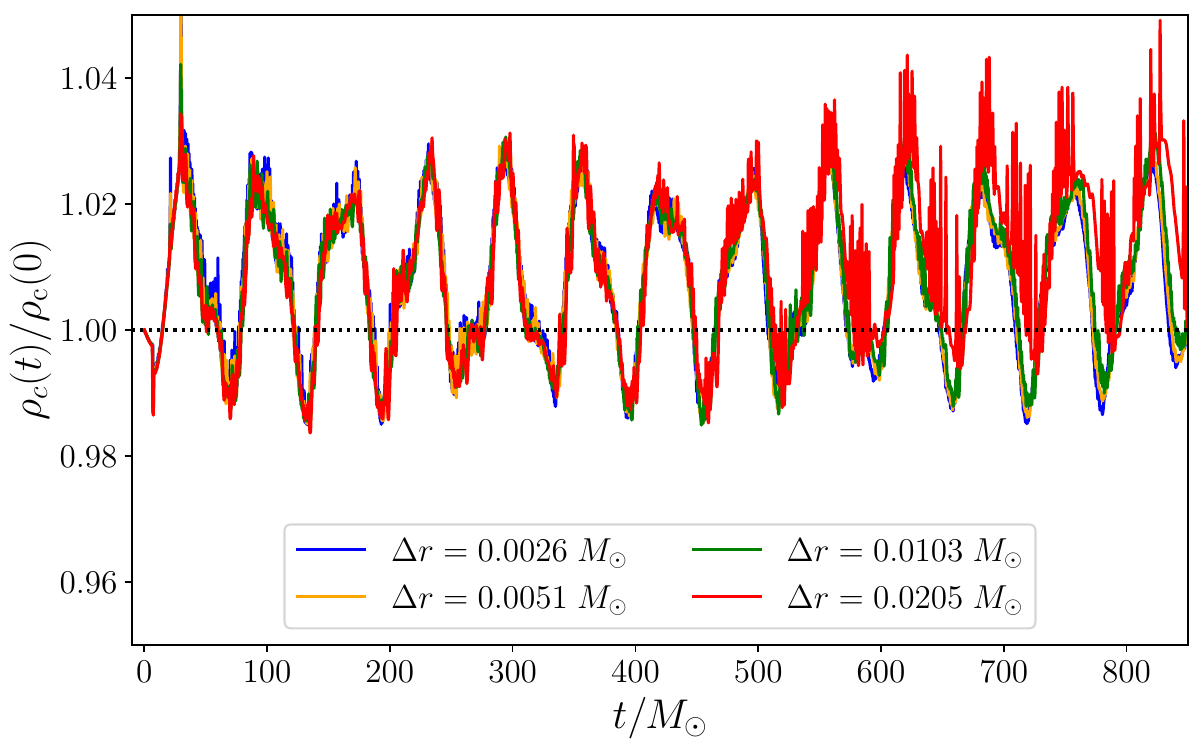}
    \includegraphics[width=0.95\linewidth]{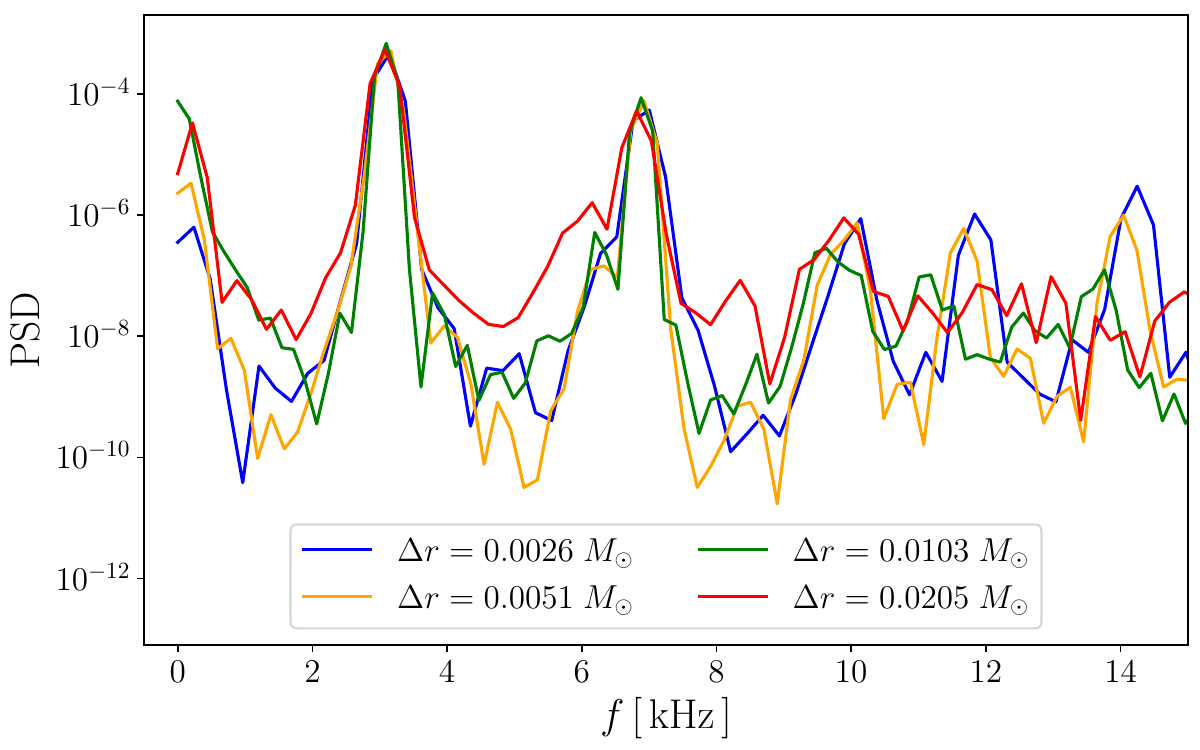}
    \includegraphics[width=0.95\linewidth]{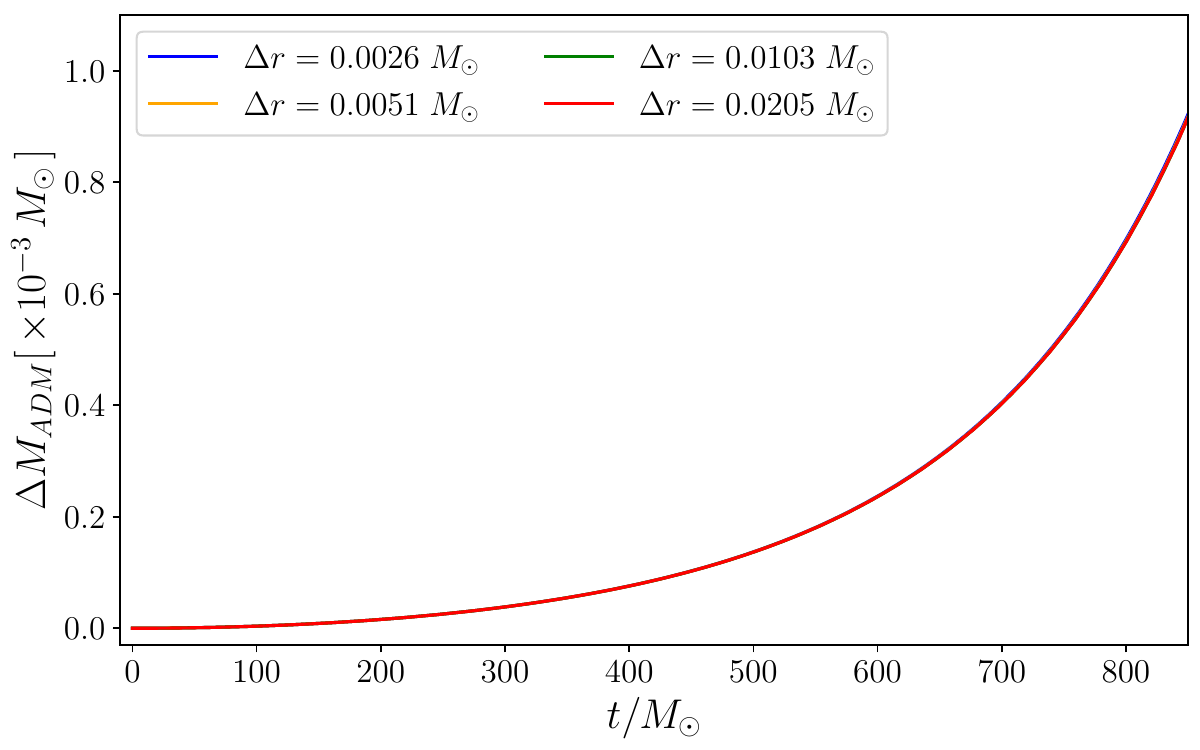}
    \caption{Upper panel: Central density time evolution. Middle panel: Power Spectrum Density.
    \textcolor{black}{Bottom: Change of the ADM mass  over time }. For all simulations, the configuration corresponds to an isotropic star with $1.6\, M_\odot$ and the QHC21 EOS}
    \label{fig:app_rho}
\end{figure}

The upper panel of Figure~\ref{fig:app_rho} shows the time evolution of the star's central density for all resolutions considered. In each case, the star exhibits stable oscillations around the initial equilibrium configuration. To quantify the fundamental frequency of the oscillations, we computed the PSD of the central density evolution for each resolution. The resulting spectra display good agreement for the first two frequency peaks across all runs, confirming the robustness of the numerical results (see middle panel of Figure~\ref{fig:app_rho}). \textcolor{black}{The frequency of the fundamental mode is $\{3.129, 3.123, 3.043,3.049\}$~kHz for $\Delta r = \{0.0026, 0.0051, 0.0103, 0.0205\}\,M_\odot$, respectively  }. As expected, higher spatial resolutions enable the resolution of additional peaks at higher frequencies.

\textcolor{black}{The bottom panel of Figure~\ref{fig:app_rho} shows the evolution of the change in the ADM mass for the different resolutions, which remains below $10^{-3} M_\odot$. We compute the ADM mass using the value of the metric component $g_{tt}$ at the outermost grid point, taking into account that, at large radii, it behaves as}
\begin{equation}
-g_{tt}=\alpha^2 \approx 1-\frac{2M_{\rm ADM}}{r}.
\end{equation}

\begin{figure}
    \centering
    \includegraphics[width=0.95\linewidth]{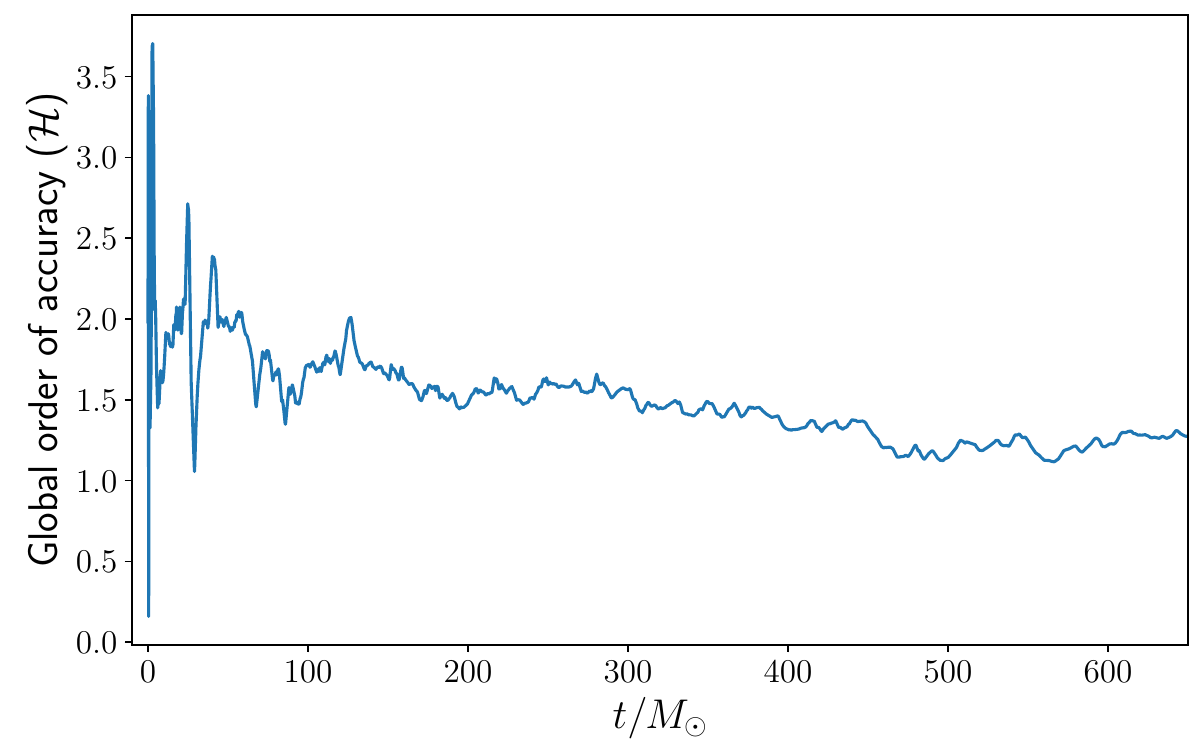}
    \caption{Global order of accuracy of the simulations. }
    \label{fig:app_p}
\end{figure}

Finally, we monitor the Hamiltonian constraint (equation~\ref{eq:Einstein1}) and compute the global order of accuracy of the code, $\tilde{p}$, following \citep{2013rehy.book.....R}: 
\begin{equation}
     \tilde{p} =  \log_2| \mathcal{R}(h, h/2, h/4)  -1 |\, ,
\end{equation}
where
\begin{equation}
  \mathcal{R}(\Delta r, \Delta r/2, \Delta r/4 ) = \frac{E^{(\Delta r)}- E^{(\Delta r/4)}}{E^{(\Delta r/2)}- E^{(\Delta r/4)}}   \, ,
\end{equation}
where
\begin{equation}
    E^{(\Delta r)}= \sum_{i} \left|\partial_r a - a^3\left[ 4\pi r (\tau + D) -\frac{m}{r^2}\right] \right|_i \Delta r \, ,
\end{equation}
with the sum extending over all grid points.
Figure~\ref{fig:app_p} shows that the estimated order of convergence oscillates between $1.5$ and $2$, consistent with the expected accuracy of the numerical methods implemented in the code.

\subsection{Neutral stability line convergence}

\begin{figure}
    \centering
    \includegraphics[width=0.95\linewidth]{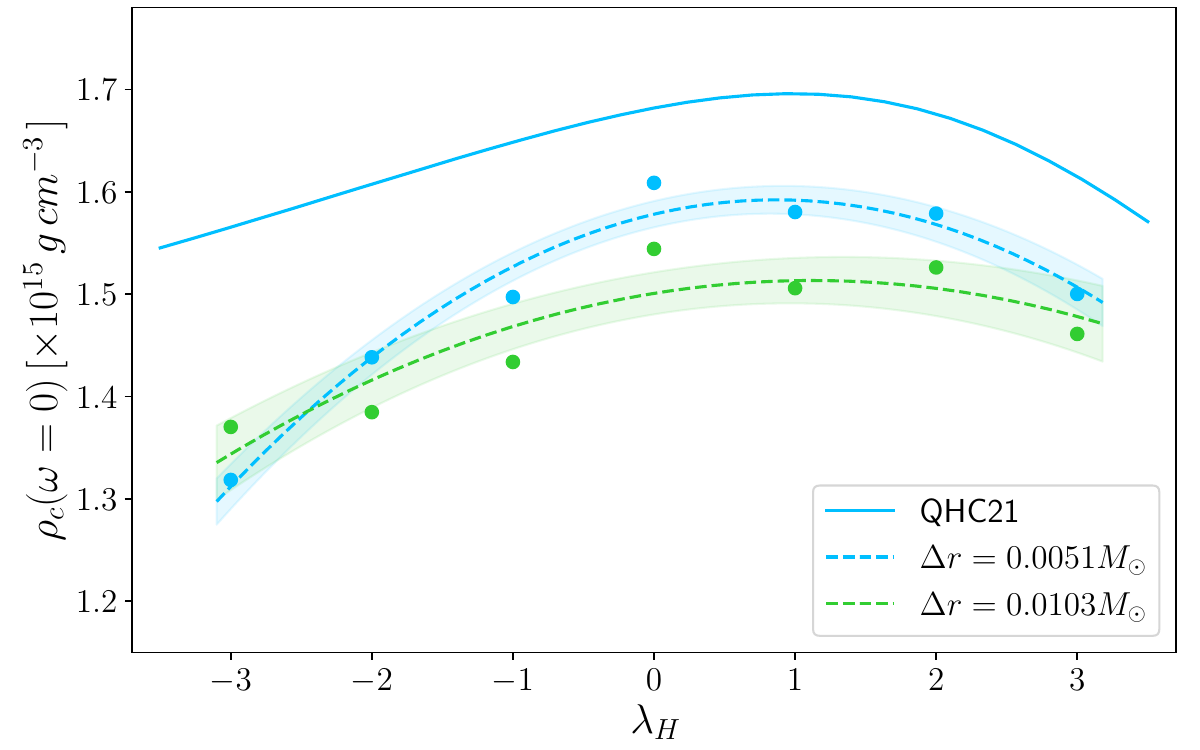}
    \caption{ \textcolor{black}{Critical density as a function of the anisotropy parameter at which the fundamental mode frequency vanishes at different resolutions for the Horvat anisotropy model and the QHC21 EOS.} }
    \label{fig:app_hovart}
\end{figure}

\textcolor{black}{We apply the same procedure described for Figure~\ref{fig:Omegaani} to the Horvat anisotropy model using the QHC21 EOS, with a radial resolution of $\Delta r = 0.0103\,M_\odot$. For a given value of the anisotropy parameter $\lambda_H$, we extrapolate the squared fundamental frequency to determine the central density at which it vanishes,  locating the neutral-stability line. Figure~\ref{fig:app_hovart} and Table~\ref{tab:nsl_hovart} compare these results across different resolutions, showing that the relative error is below 5\%.}

\begin{table}[h!]
    \centering
    \begin{tabular}{c|ccc}
    \hline
    &  & & \\
     & \multicolumn{3}{c}{$\rho_c(\omega=0) \times 10^{15}$ g cm$^{-3}$}\\ 
       $\lambda_H$  & $\Delta r= 0.0051M_\odot$& $\Delta r= 0.0103 M_\odot$ &  error\\ 
        &  & & \\ \hline 
         &  & & \\ 
         $-3$ & $1.318$ & $1.370$ & $0.039$ \\
         $-2$ & $1.438$ & $1.384$ &$0.037$ \\
         $-1$ &$1.497$ & $1.433$ & $0.042$\\
         $0$ & $1.608$ & $1.544$ &$0.040$ \\
         $1$ & $1.580$ & $1.505$ & $0.047$\\
         $2$ & $1.578$ & $1.526$ &$0.033$ \\
         $3$ & $1.500$ & $1.461$ & $0.026$\\ 
          &  & & \\ \hline\hline
    \end{tabular}
    \caption{\textcolor{black}{Comparison of the critical central density at which the fundamental-mode frequency vanishes, for the Horvat anisotropy model and the QHC21 EOS, computed at different numerical resolutions.}}
    \label{tab:nsl_hovart}
\end{table}

\section{Radial perturbation}\label{app:II}

To study radial perturbations in equilibrium configurations, one considers small, time-dependent deviations from a static, spherically symmetric background. These perturbations are applied to both the metric and the fluid variables, $X$, typically expressed in Eulerian form as: , where $X_0(r)$ denotes the background (unperturbed) solution and $\delta X_0(r)$  represents the amplitude of the perturbation.

As a result of the perturbation, the fluid elements undergo motion: a fluid element initially located at radial coordinate $r$ in the unperturbed configuration is displaced to a new position $r+\xi(r,t)$, where $\xi=\xi(r)e^{-i\omega t}$  is the Lagrangian displacement. Expanding the Einstein field equations to linear order in the Lagrangian displacement yields the following set of perturbation equations \cite{2020EPJC...80..726P}:
\begin{eqnarray}\label{eq:dchidr}
    r\frac{d \chi}{dr}&=&\frac{-\Delta P}{c_s^2 (P_0+ \epsilon_0)}  + \chi\left(\frac{2\sigma_0}{P_0+\epsilon_0} + \frac{r}{\alpha_0}\frac{d\alpha_0}{dr}-3  \right)
 \end{eqnarray}
    
\begin{eqnarray}
       \frac{d \Delta P}{dr}&=& \left(-\frac{4\pi r^2(P_0+\epsilon_0)}{r-2m_0}-\frac{1}{\alpha_0}\frac{d\alpha_0}{dr}\right)\Delta P +  \label{eq:DdeltaP}\\
      & & 2(b_1(P_0+\epsilon_0)-\sigma_0) \frac{d\chi}{dr} + (P_0+\epsilon_0)  \left[  \bar\sigma + \right . \nonumber\\
      & & \frac{r^2\omega^2}{\alpha_0^2(r-2m_0)} - \frac{8\pi r^2P_0}{r-2m_0} + \nonumber \\
      & & \left. \frac{2}{\alpha_0}\left(2-\frac{\sigma_0}{P_0+\epsilon_0}\right)\frac{d\alpha_0}{dr}+ \frac{r}{\alpha_0^2}\left(\frac{d\alpha_0}{dr}\right)^2 \right]\chi \nonumber
\end{eqnarray}
with 
$$\bar \sigma =b_1\left(-\frac{6}{r}-\frac{2}{P_0+\epsilon_0}\frac{dP_0}{dr}\right) + \frac{b_2}{\alpha_0}\frac{\alpha_0}{dr}$$

where $\chi=\cfrac{\xi}{r}$, $c_s^2=\cfrac{dP}{d\epsilon}$ , $b_1=\cfrac{d\sigma}{d\epsilon}$ and $b_2=\cfrac{d\sigma}{d\mu}$ with $\mu=2m_0/r$, and $\Delta P$ is the lagrarian perturbation of the pressure. To derive the above equations, a barotropic EOS for the radial pressure has been assumed, i.e. $P=P(\epsilon)$, and that the anisotropy has the following dependence: $\sigma=\sigma(\epsilon, \mu)$

Equations~(\ref{eq:dchidr}) and (\ref{eq:DdeltaP}) are integrated using the following initial conditions at the center of the star ($r\rightarrow 0$):
$$\xi\rightarrow0 \quad ;\quad \chi\rightarrow1 \quad;\quad   \Delta P \rightarrow  -3 ( P_0 + \epsilon_0 )   c_s^2  \, , $$
and the boundary condition at the stellar surface ($r\rightarrow R_{\rm star}$):
$$ \Delta P\rightarrow 0\, .$$
For a given stellar configuration, equations (\ref{eq:dchidr}) and (\ref{eq:DdeltaP}) yield a discrete set of eigenvalues $\omega_n^2$, each with associated eigenfunctions $\xi_n(r)$ and $\Delta P_n(r)$. The lowest eigenvalue represents the fundamental radial mode. We numerically determine these $\omega^2$ values using a shooting method: we integrate the system for trial values of $\omega^2$ and identify the correct eigenfrequencies as those that satisfy the boundary condition at the stellar surface.

\begin{acknowledgments}
\textcolor{black}{We thank the anonymous referee for their valuable comments and suggestions. }
 E.~A.~B-V is supported by the Vicerrector\'ia de Investigaci\'on y Extensi\'on - Universidad Industrial de Santander Postdoctoral Fellowship Program No. 2025000167. J.~F.~R is supported by the Vicerrector\'ia de Investigaci\'on y Extensi\'on - Universidad Industrial de Santander Postdoctoral Fellowship Program No. 2025000174. 
\end{acknowledgments}


\bibliography{apssamp}

\end{document}